\documentclass[preprint2]{aastex}

\usepackage{amsmath}
\usepackage{graphicx}
\usepackage{natbib}
\usepackage{txfonts}
\usepackage{amssymb}
\usepackage{rotating}

\bibpunct{(}{)}{;}{a}{}{,}
\newcommand{\myarcsec}{\hbox{$.\!\!^{\prime\prime}$}}

\DeclareRobustCommand{\ion}[2]{\textup{#1\,\textsc{\lowercase{#2}}}}

\shorttitle{Optical, near- and mid-IR imaging reduction with THELI}
\shortauthors{Mischa Schirmer}

\begin{document}
\bibliographystyle{aa}

\title{THELI -- Convenient reduction of optical, near- and mid-infrared
     imaging data}


\author{M. Schirmer\altaffilmark{1,2,3}}
\affil{\altaffilmark{1}Gemini Observatory, Casilla 603, La Serena, Chile; \email{mschirme@gemini.edu}}
\affil{\altaffilmark{2}Isaac Newton Group of Telescopes, Santa Cruz de La Palma, Spain}
\affil{\altaffilmark{3}Argelander-Institut f\"ur Astronomie, Universit\"at Bonn, Germany}
 
\begin{abstract}
The last 15 years have seen a surge of new multi-chip optical and near-IR imagers. While some of 
them are accompanied by specific reduction pipelines, user-friendly and generic reduction tools 
are uncommon. In this paper I introduce THELI, an easy-to-use graphical interface driving an 
end-to-end pipeline for the reduction of any optical, near-IR and mid-IR imaging data. The advantages of 
THELI when compared to other approaches are highlighted. Combining a multitude of processing algorithms 
and third party software, THELI provides researchers with a single, homogeneous tool. A short learning 
curve ensures quick success for new and more experienced observers alike. All tasks are largely automated, 
while at the same time a high level of flexibility and alternative reduction schemes ensure that widely 
different scientific requirements can be met.
Over 90 optical and infrared instruments at observatories world-wide are pre-configured, 
while more can be added by the user. The online Appendices contain three walk-through examples using
public data (optical, near-IR and mid-IR). Additional extensive online documentation is available 
for training and troubleshooting.
\end{abstract}


\section{Introduction}
The systematic introduction of CCDs revolutionized the field of observational astronomy at the beginning 
of the 1980s \citep[see e.g.][]{crs81,gob81,guw81,mch81,mof81}. It was paralleled by the development of 
large software packages to process these new kinds of data, such as IRAF \citep{bus81,val84}, ESO-MIDAS 
\citep{bcg83}, and STARLINK \citep{paw80}. These programs have become deeply tied to observational 
astronomy. They form an integral part of the observing and data acquisition software at many 
observatories, and numerous reduction packages have been built on them.

At the end of the 1990s the limited field of view of a single CCD was overcome by mosaic cameras, such 
as (in the optical) WFI at the Anglo-Australian Observatory (AAO), WFI at the 2.2m MPG/ESO telescope
\citep{bmi99}, CFH12K at the Canada-Hawaii-France Telescope \citep[CFHT,][]{cls00}, and SuprimeCam at 
Subaru \citep{mks02}, and in the near-IR CIRSI at the William Herschel Telescope \citep[WHT,][]{bmm97}. 
In the following, these and other instrument / telescope combinations will be abbreviated as in e.g. 
WFI@AAO. Deep and wide surveys have 
become possible since then, drastically increasing the volume of data produced per night. While the new 
instruments have been 
used frequently from the start, the volumes and much increased complexity of the data have proved a 
challenge for the first few years as corresponding software support lagged behind. In particular 
astrometric and photometric solutions have become non-trivial and tiresome, now that dithered images 
from detector mosaics with distortion patterns have to be combined. This has triggered the development 
of many astrometric pipelines outside the IRAF or MIDAS context, such as {\tt WIFIX}\footnote{M. Radovich, 
{\tt http://www.na.astro.it/$\sim$radovich/wifix.htm}}, the {\tt LDAC} pipeline\footnote{E. Deul, 
{\tt ftp://ftp.strw.leidenuniv.nl/pub/ldac/software/}}, and {\tt Scamp} \citep{ber06}, to name just a few.

New survey telescopes such as the VST, VISTA, and the future LSST are pushing the boundaries further, 
combining extreme optics with detector mosaics of up to a hundred CCDs or more (e.g.~DECam@CTIO, 
\cite{faa12}; GPC1@Pan-Starrs, \cite{oti08}; HyperSuprimeCam@Subaru, \cite{mkn12}; ODI@WIYN, \cite{jtb02};
OmegaCam@VST, \cite{kpc02}; VIRCAM@\-VISTA, \cite{hhb10}).
While some of these cameras are mostly used for public surveys, others are facility instruments 
available for private investigators (PIs). The associated data reduction pipelines are not necessarily 
publicly available (e.g.~VISTA) or portable to the PI's computer, as they may be tied to large 
commercial data base applications.

Data processing is made even more challenging by the multitude of optical and near-IR data we
combine for our research. It is common that exposures from several telescopes and 
cameras are pooled for one project, providing sufficient sample sizes and/or a multi-wavelength 
perspective. Frequently, researchers have to switch to new and unfamiliar software to ensure consistent 
data quality, a time-consuming and demanding task. Custom-made reduction scripts and methods 
are often no longer interchangable between instruments.

In 2000, Thomas Erben and I faced these problems for our 20 square degree weak lensing survey 
\citep{ses03}, and decided to develop an instrument-independent pipeline that could process data from 
any future, optical multi-chip camera. We exploited existing, stand-alone software solutions (mostly
{\tt Astromatic}\footnote{{\tt http://www.astromatic.net}} and {\tt LDAC}), and merged them with 
custom-made modules to ensure a fast and smooth data flow. This THELI pipeline core is operated by 
a number of command-line scripts, and has been developed on data from WFI@2.2m MPG/ESO and 
MegaCam@CFHT. Its main purpose is the reduction of large survey data sets, such as 
the CFHTLenS \citep{ehm13}, utilizing Linux cluster architectures. The pipeline, presented in 
\cite{esd05} (hereafter E05), has been extensively tested \citep[e.g.][]{hed06,ehm13}.

The command-line approach for mass production competed with the goal for an instrument-independent pipeline 
to be used by a PI for his or her individual projects. Hard-coded parameters and manual editing of configuration 
files were just some of the obstacles. Another problem is that the command-line scripts have been optimised for 
the reduction of blank fields, and do not work (well) for crowded fields or significantly extended sources. To 
overcome these issues I started developing a THELI graphical user interface (GUI). It includes many new tasks such 
as alternative astrometric routines and more adaptive sky subtraction methods. In addition to full near- and mid-IR 
support it also handles all administrative tasks for the user. Thus, a highly flexible and convenient tool able to 
process essentially all kinds of imaging data now exists. It facilitates a broad range of scientific 
studies, ranging e.g.~from the optical distortion modeling of the LORRI camera on-board the \textit{New Horizons} 
mission \citep{mmb12} or the \textit{Deep Impact} campaign \citep{maa05}, via the expansion history of planetary 
nebulae \citep{scm08} to studies of galaxy clusters \citep{llh12}. THELI's command-line version has been used for 
several large survey projects, such as a systematic lensing study of 51 of the most X-ray luminous galaxy clusters
known \citep{laa12}, and the CFHT Lensing Survey \citep[CFHTLenS, comprising 154 square degrees;][]{hwm12}.

This paper contains a description of the THELI GUI's (version 2.8.1) most important features. 
An additional comprehensive user manual and technical reference is available 
online\footnote{{\tt http://www.astro.uni-bonn.de/$\sim$theli/gui}}. For the remainder of this 
paper, `THELI' refers to the GUI. Section 2 contains a short overview 
of the working principle, and discusses software pre-requisites for installation. Section 3 covers the 
preparation of the data, non-linearity and cross-talk corrections. In Section 4 the various
options available for background modeling are explained. This is followed by a short summary about 
weighting and bad pixel mapping in Sect.~5, with astrometric and photometric calibration in Sect.~6. Sky 
modeling is addressed in Sect.~7, and the various options available for final image coaddition in Sect.~8. 
A short summary and outlook is presented in Sect.~9. Appendices A through C (online material) 
contain three step-by-step reduction examples, based on public optical, near-IR, and mid-IR images.
A list of currently supported instruments is included at the end.

\begin{figure*}[t]
  \includegraphics[width=1.0\hsize]{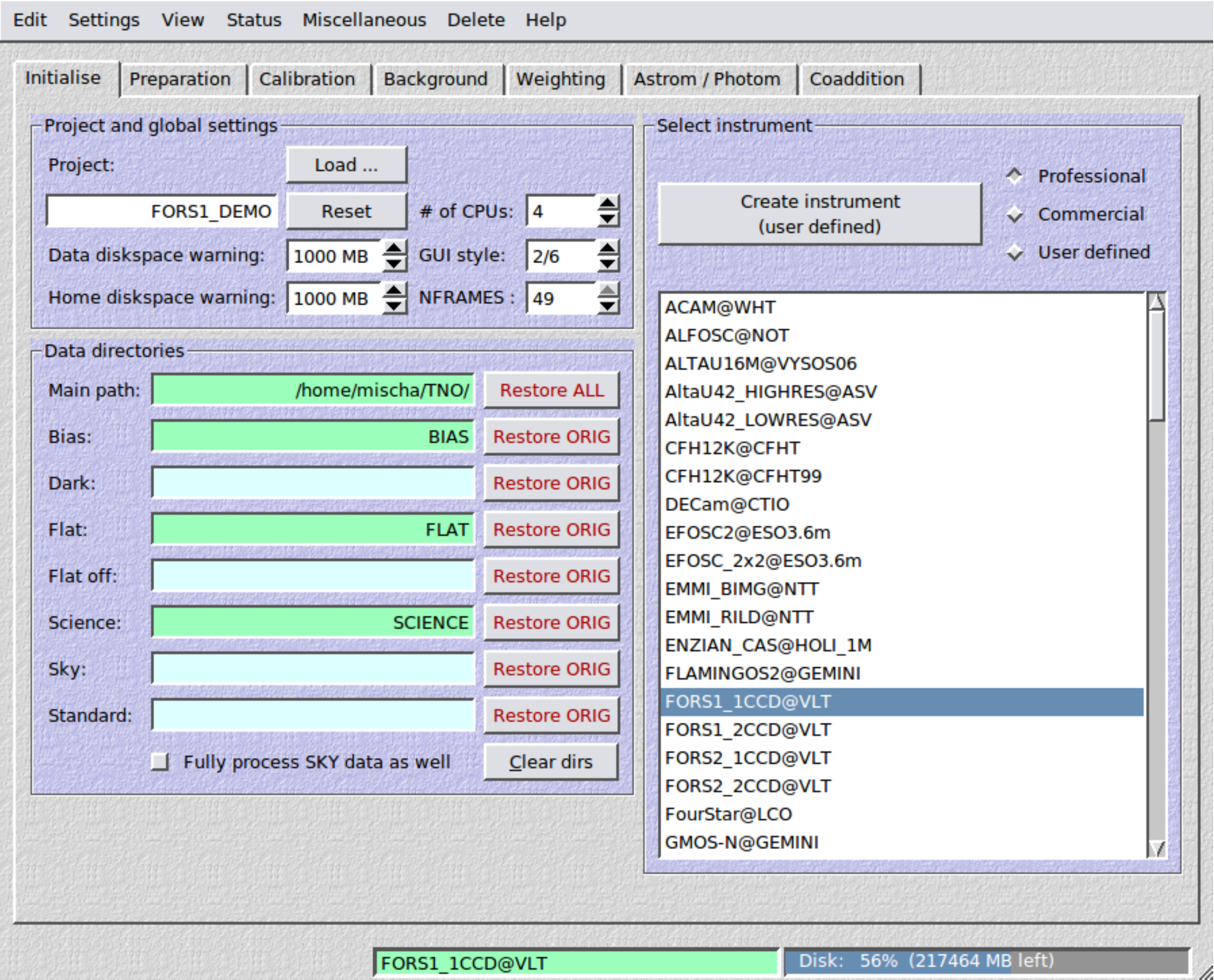}
  \caption{\label{theliini}{Initial setup of a new data reduction project in THELI.}}
\end{figure*}

\section{Structure and operation of THELI}
Most data processing algorithms in THELI are not new inventions, but have been used before.
The big advantage of THELI is that a multitude of these procedures has been combined and made accessible 
in a single, homogeneous tool. Figure \ref{theliini} shows the \textit{Initialise} section of THELI's 
user interface, where global information is provided, such as the current project title, how many CPUs 
to use, where the data can be found, and the instrument they were taken with. Thereafter, the user steps 
through six sections to arrive at a coadded image. The main aspects of these sections are discussed in 
this paper, with concrete examples given in the appendices.

\subsection{Software pre-requisites and implementation}
Most software modules developed for THELI, as well as third party programs such as the 
{\tt Astromatic} tools, are written in C/C++. They are compiled with standard tools such as {\tt gcc} 
and {\tt make}, and do not require exotic libraries. A few other modules are based on the {\tt python} 
language. The user interface itself is based on the C/C++ {\tt Qt} library, which is also the heart of many 
Desktop environments such as KDE. All these pre-requisites are fundamental parts of modern Linux systems 
and available in pre-compiled form in their respective repositories. As such, THELI can be compiled on 
essentially all Linux platforms and has sufficient backwards compatibility.

The only current shortcoming is that THELI does not compile under MacOS\footnote{The command-line version, 
however, using the pipeline core only, does compile under MacOS.}. This can be overcome by 
installing a Linux guest operating system in a virtual machine, at the cost of performance. THELI is 
currently based on {\tt Qt3}, which is being phased out in future Linux releases, but can always be compiled 
from source\footnote{{\tt http://download.qt-project.org/archive/qt/3/}}. In the near future, THELI will be 
ported to the recently released {\tt Qt5} library, and offer full support for MacOS.

\section{\label{splitting}Correcting headers and detector features}
\subsection{\label{fixheader}Homogenizing FITS headers}
The first step in THELI is to bring all FITS headers from different instruments to a common 
format, taking into account possible header changes made over time. This is essential 
for an instrument-independent pipeline to work, as not all FITS headers adhere to the FITS 
standard\footnote{{\tt http://fits.gsfc.nasa.gov/fits\_standard.html}}. For example, conflicting 
keywords such as {\tt CDij} and {\tt PCij} may appear simultaneously, or the instrument's orientation 
on sky is unknown. Likewise, the WCS information may be inaccurate or have to be reconstructed from 
nonstandard keywords. Invalid key values such as negative airmass have to be corrected as well. 
More intricate problems such as incorrect filter names can in general not be detected automatically. 
Thus it is the user's responsibility to validate the integrity of the raw data.

Consequently, THELI modifies the original FITS header, retaining and translating only keywords essential 
for the data reduction. Apart from mandatory keywords, the new header comprises a complete set of WCS 
keys, date and time, plus a few others such as airmass, filter and exposure time. Additional keywords to be 
retained can be defined by the user. Multi-extension FITS files are split into individual chips for 
parallelisation purposes at this point of the processing.

\subsection{Cross-talk}
\begin{figure}[t]
  \includegraphics[width=1.0\hsize]{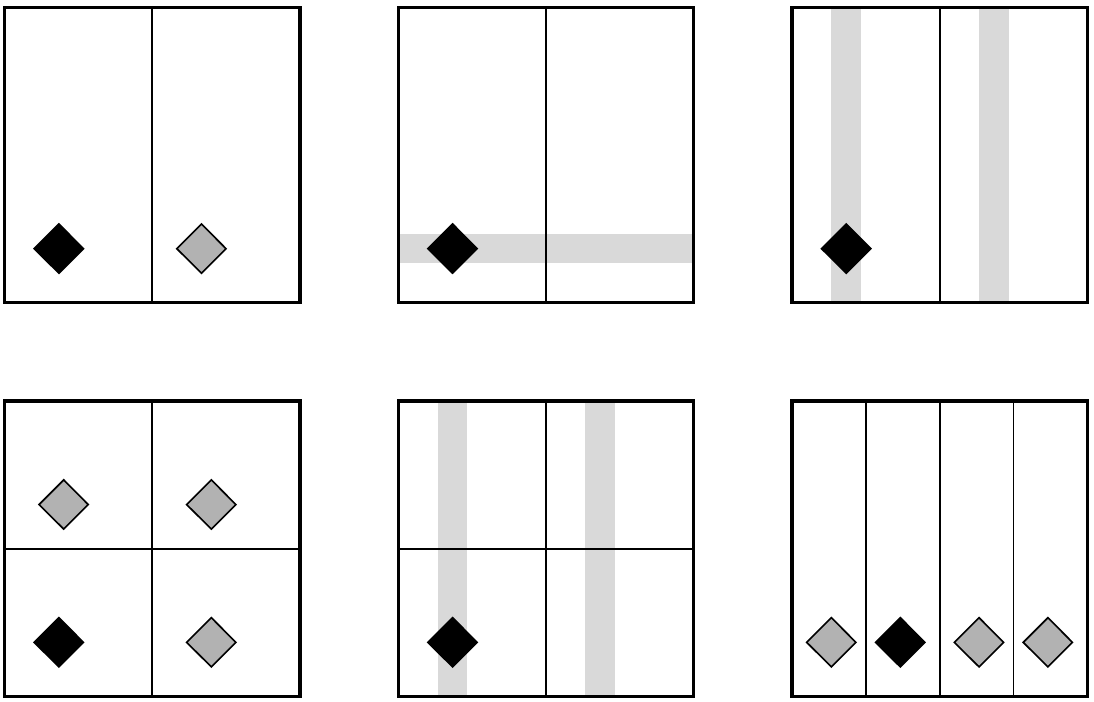}
  \caption{\label{xtalk}{Various cross-talk effects. The offending source and its ghosts from 
      normal cross-talk are shown as black and gray diamonds, respectively. Shaded gray lines 
      represent row cross-talk. The thin lines mark the readout sections. Both normal and row 
      cross-talk can appear simultaneously.}}
\end{figure}

Different forms of cross-talk (Fig.~\ref{xtalk}) may appear in detectors with multiple readout 
channels. \textit{Intra-chip normal cross-talk} causes ghost images of a bright source in the other 
channels. \textit{Row cross-talk} enhances the values of the rows (or columns) containing
a bright source, and the same feature is projected into the other readouts \citep[for an example see][]{faa01}. 
These effects are fixed by subtracting a rescaled image (or scaled average row values) of the offending 
channel from the other channels. Both forms of cross-talk may appear simultaneously.

\textit{Inter-chip cross-talk} can occur in multi-chip cameras, e.g.~WFC@INT, DECam@CTIO and OmegaCam @VST,
creating ghost image of one chip in another chip. It can be corrected by image 
subtraction, a mode that is currently being implemented in THELI.

Some HAWAII 2048$\times$2048 arrays show \textit{edge cross-talk}, recognized by positive and 
negative amplitudes within a ring-shaped ghost. \cite{abw08} show that this can be solved 
for on a hardware level. THELI can in principle also correct for this effect. However, results have 
not been satisfactory for the cameras tested (HAWK-I@VLT, VIRCAM@VISTA, Flamingos2@Gemini), as 
the amplitudes of the ghosts vary between channels.

In general, cross-talk amplitudes may be positive or negative, i.e.~the ghost images can be bright 
or dark. Their amplitudes are typically a factor $10^{4}$ smaller than the sources, but may become 
as large as $0.025$, e.g.~for NICS@TNG. Corrections work best if the offending sources are not saturated. 
Cross-talk severely compromises the data from crowded fields and deep observations, as the ghosts are fixed 
with respect to the sources and thus do not appear as outliers in a set of dithered images. By 
choosing different sky position angles, however, cross-talk can be statistically identified and removed.

\subsection{Non-linearity}
THELI's non-linearity correction is represented by a third-order polynomial, 
\begin{equation}
x_{\rm lin} = \sum\limits_{i=0}^3 a_i\,{x_0}^i,
\end{equation}
where $x_0$ are the pixel values in the uncorrected exposure. The coefficients $(a_i)$ are stored 
in a separate configuration file. Currently, only a few instruments (such as WFC@INT) are pre-configured 
for non-linearity correction. Coefficients for other instruments can be added by the user.

After these preparatory steps, the user creates a master bias and a master flat, and applies them to 
the data. This is a straightforward process and described in the example reductions in the appendices.

\begin{figure}[t]
  \includegraphics[width=1.0\hsize]{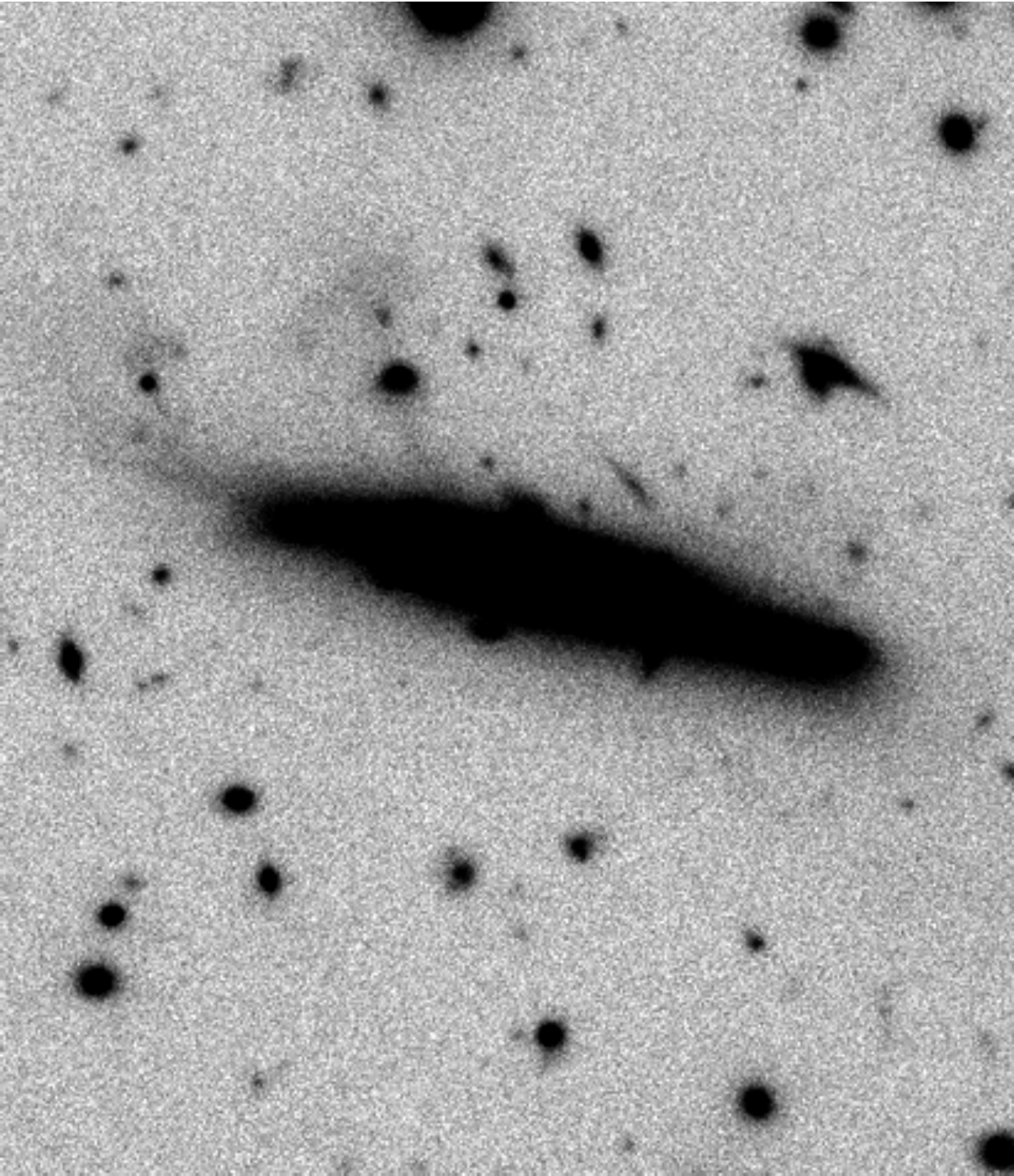}
  \caption{\label{anon}{Deep HAWK-I $J$-band image of an anonymous field galaxy. A faint tidal plume 
      and other debris from minor mergers are clearly visible, demonstrating the quality of THELI's 
      near-IR background modeling.}}
\end{figure}

\section{\label{background}Background modeling and zero-point variations}
THELI distinguishes between \textit{background modeling} and \textit{sky subtraction}. A background model is 
a median combination of several successive and dithered exposures, correcting for residual variations 
after flat-fielding. Depending on the cause of the variations, the images must be divided (`superflatted') 
by the model, or the model is subtracted. The latter is typically the case for near-IR observations 
and removes most of the sky signal. Sect.~\ref{staticdynamic} and \ref{skydata} show different ways 
of creating a background model. \textit{Sky subtraction}, on the other hand, is based on 
single images and eliminates any remaining individual gradients or pedestals (see Sect.~\ref{sky}).

Background variations may vary over time. Possible causes include both additive and multiplicative effects, 
such as
\begin{itemize}
\item{improper illumination of the domeflat screen,}
\item{scattered light in twilight flats,}
\item{moonlight,}
\item{airglow (particularly in the near-IR),}
\item{sky concentration (reflection between CCD and corrector lenses or filters),}
\item{fringing.}
\end{itemize}

Spatial zero-point variations occur when additive and multiplicative components are mixed in 
a flat field. They result either from external scattering (e.g.~unfavorable dome orientation during 
twilight), or internal reflections in the instrument. The amplitudes of these variations can 
reach up to 10\%. A direct comparison of stellar magnitudes in dithered data is required. A comparison 
with known reference sources in the field can also be performed to solve this problem \citep{mas01,kog04}. 
The latter approach is available in THELI (Sect.~\ref{directphot}), yielding single zero-point 
adjustments for individual chips but not yet a full 2D correction. For instruments that are mostly used 
for survey work, the corresponding 2D corrections are usually included in a dedicated pipeline 
(e.g.~VIRCAM@VISTA), or directly applied to pre-processed archival data \citep[e.g.~MegaCam@CFHT, using 
the ELIXIR pipeline;][]{mac04}.

\subsection{\label{staticdynamic}Static and dynamic background models}
A \textit{static background model} is a single median image created from a series of exposures. 
It can be applied to all images from which it was created, and is rescaled to correct for 
slow, global variations. A good example is a background model created from a $10-30$ minute long 
sequence of short near-IR exposures. Atmospheric airglow changes little during this time, and thus
a single static model is sufficient to subtract the sky. Another case would be a classic optical 
superflat computed from data taken on a clear night.

A \textit{dynamic background model} is needed if sky conditions are unstable during a series of exposures.
This is generally the case for any near-IR sequence extending over more than $10-30$ minutes, as the 
airglow changes the sky brightness on time scales of minutes with spatial fluctuations on scales of 
arcminutes. Optical $i$- or $z$-band data may also require dynamic correction. THELI creates the background 
model for the $k$-th exposure from the $m$ closest exposures in time. The latter may or may not bracket the 
$k$-th exposure symmetrically.

Common to both approaches is that all objects are masked prior to the creation of the model,
and the highest pixel in the stack is rejected. These settings can be adjusted arbitrarily. THELI 
also supports two-pass background subtraction, which may be required in the near-IR for extended objects 
or very deep observations (see App.~\ref{example2} for an example processing flow). The quality of 
THELI's background models is illustrated in Fig.~\ref{anon}, revealing two extended tidal debris 
features of low surface brightness around a random field galaxy \citep[see also][]{mgc10}.

An exposure sequence containing one or more longer interruptions is not adequately represented by a single
background model. If the interruptions exceed a user-specified amount of time, separate background models 
are automatically calculated for each group of images and applied accordingly.

\subsubsection{\label{skydata}Separate sky observations}
The variable near-IR background requires larger dither patterns in the presence of extended objects
to be able to estimate the sky at the object's position. A nearby blank field has to be observed in a 
periodic manner if the target comprises about half the detector's field-of-view or more. This may also 
become necessary for optical observations if an accurate correction is required. Exposures of a blank 
area should be kept in a separate {\tt SKY} directory where they are found and processed by THELI.

\begin{figure}[t]
  \includegraphics[width=1.0\hsize]{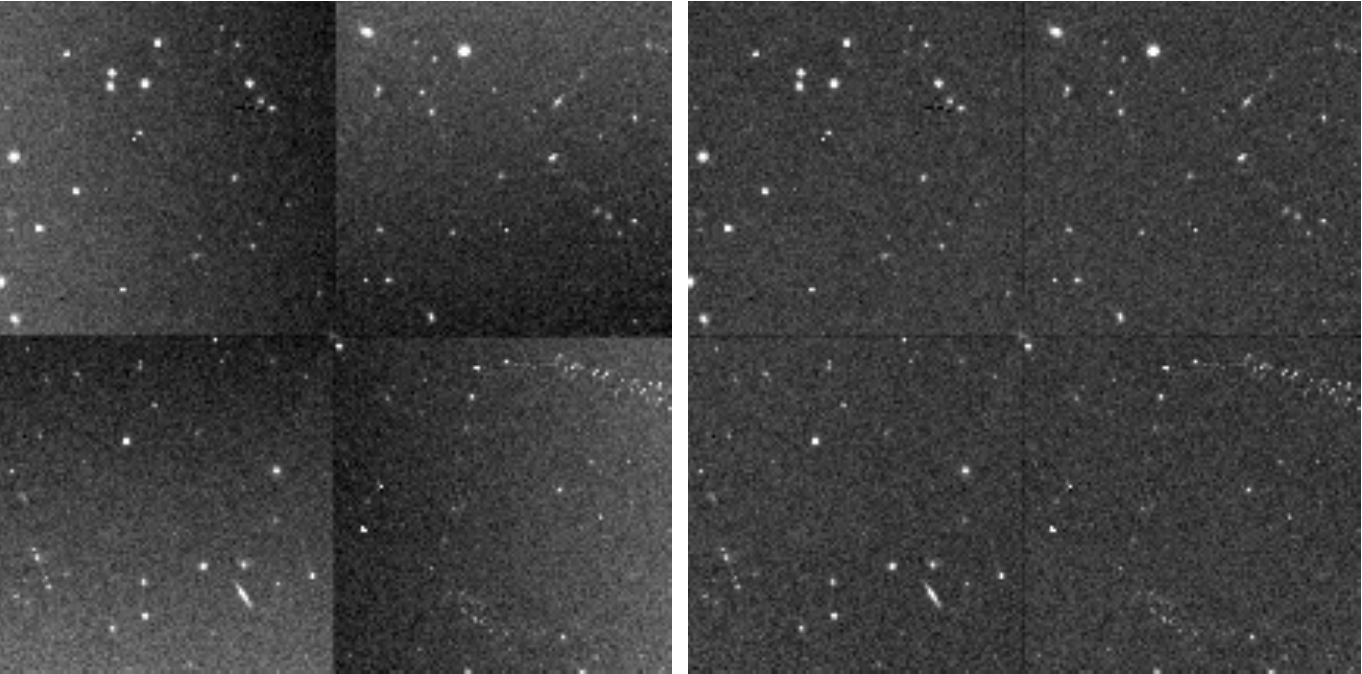}
  \caption{\label{moircs}{Left: Single MOIRCS@SUBARU $J$-band exposure after background modeling, 
      showing a reset anomaly in the four quadrants \citep[data from][]{uyk12}. Right: After 
      correction.}}
\end{figure}

\subsection{\label{collapse}Removing linear gradients}
Some HAWAII arrays show residual linear gradients after background modeling, leading to discontinuities
between readout quadrants. This \textit{detector reset anomaly} may depend on exposure time, sky brightness, 
the time passed between exposures, detector temperature, etc. It is corrected by subtracting an average 
row or column calculated from the affected area. This \textit{collapse correction} 
also supports HAWAII-2 arrays with 4 quadrants and alternating 90 degree orientation (see Fig.~\ref{moircs}).

\begin{figure*}[t]
  \includegraphics[width=1.0\hsize]{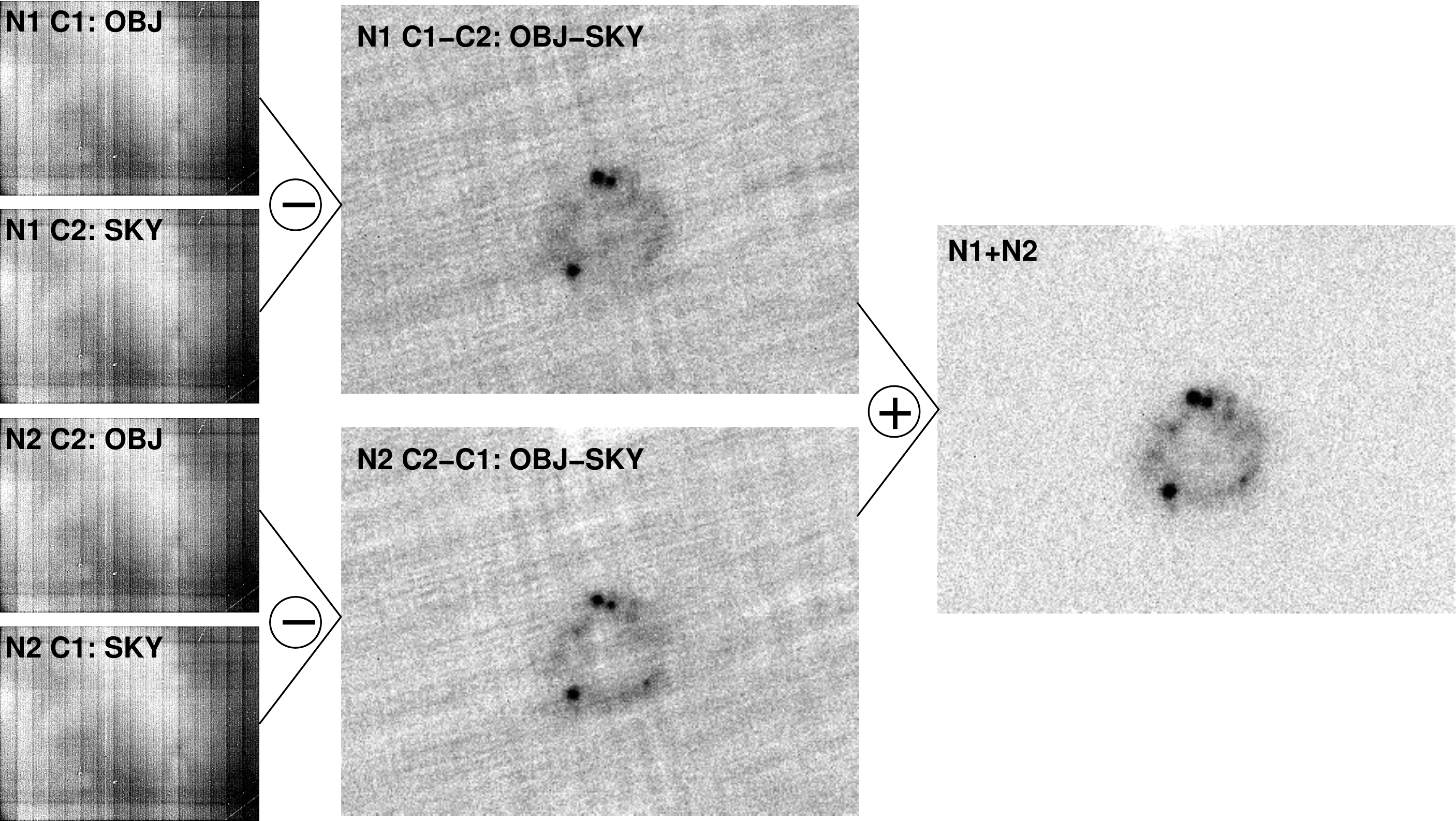}
  \caption{\label{trecs-chopnod}{Chop-nod observations of the starburst ring in NGC 7552, observed with 
      T-ReCS@GEMINI in the [\ion{Si}{II}] 8.8$\mu$m bandpass. {\tt N1/2} and {\tt C1/2} refer to the nod 
      and chop positions. With T-ReCS, the science target is observed at C2 instead of C1 when at nod 
      position 2. The residual telescopic thermal background visible in the difference images (middle column) 
      is therefore inverted. The sum of the difference images hence yields the corrected image (right).}}
\end{figure*}

\subsection{\label{midir}Mid-IR observations}
Ground-based mid-IR observations require special observing techniques as the background signal is
very high ($-5$ to $-7$ mag/arcsec$^2$). To remove the rapidly varying sky, the secondary mirror is 
tilted (\textit{chopped}) every few seconds, projecting an empty sky area in the immediate vicinity 
of the target onto the detector. The second image is then subtracted from the first one. But the thermal 
signal from the telescope does not cancel out entirely, since the detector sees the telescope 
from slightly different angles at the two chopper positions. The telescope is therefore offset 
slightly (a process known as \textit{nodding}), and another series of chopped images taken at the 
new position. A new difference image is calculated from the chopped images at the second nod position.
Finally, by subtracting it from the first difference image, the residual thermal signal is removed. 
If the order of science and sky observations are reversed at the 2nd nod position, the difference images 
have to be added (see Fig.~\ref{trecs-chopnod}). As the telescope's temperature changes only slowly, the 
nodding is usually done once or twice per minute.

Data formats and the amount of information stored in the raw data vary greatly between mid-IR instruments.
T-ReCS@GEMINI, for example, averages the two-dimensional individual exposures obtained at a given chop/nod 
position and sorts them into a 4-dimensional hyper-cube, where the two extra dimensions specify (1) the 
telescope's nod position, and (2) whether the chopper observed the sky or the target. Sky exposures with T-ReCS 
are unguided and generally discarded after processing due to inferior quality. A typical chop-nod cycle lasts 
several minutes, and is stored as one hyper-cube. An arbitrary number of such hyper-cubes may be contained in 
separate FITS extensions. THELI performs the chop-nod sky subtraction during initial data preparation.
It writes out either a single 2D FITS image, representing the fully stacked hyper-cube, or one image for every 
chop-nod cycle.

\begin{figure}[t]
  \includegraphics[width=1.0\hsize]{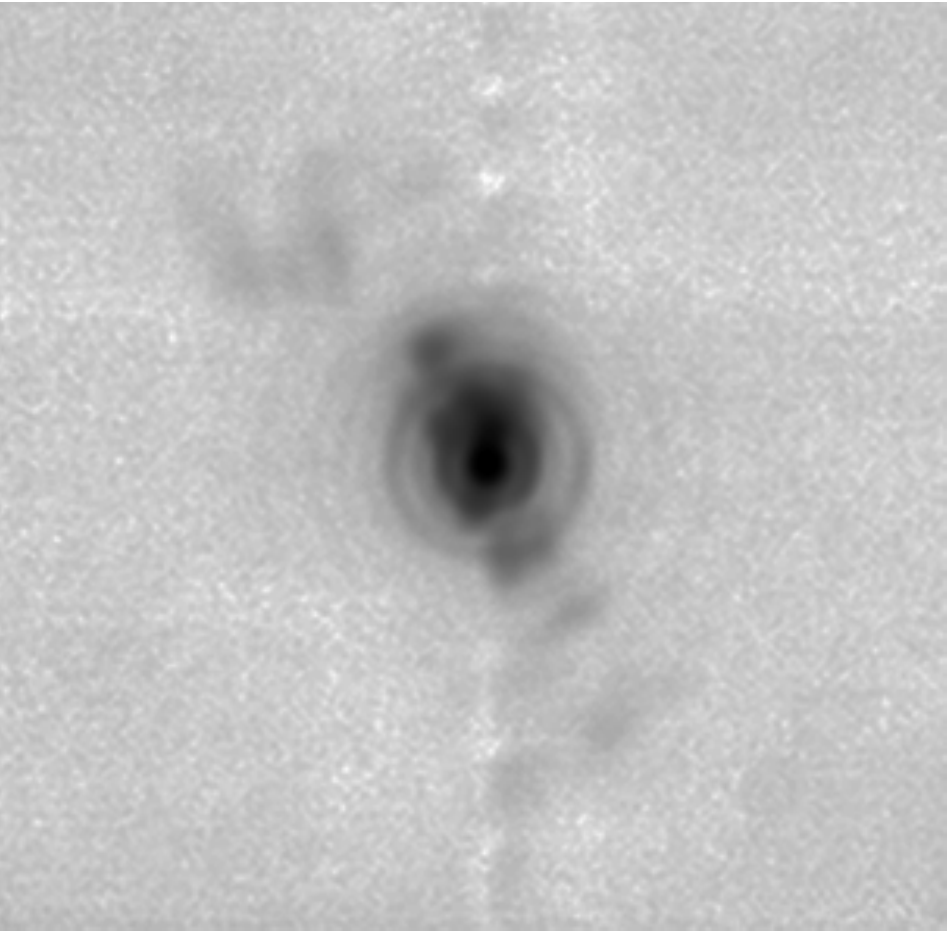}
  \caption{\label{ngc1068}{The center of the Seyfert 2 galaxy NGC 1068, as seen in [NeII] 12.8$\mu$m with 
      VISIR@VLT. Note the circularly symmetric diffraction rings.}}
\end{figure}

The other currently supported mid-IR camera is VISIR@VLT. Currently a generic chop-nod module is 
available for thermal background removal. This will be replaced by individually optimized routines (such 
as for T-ReCS) once more mid-IR cameras are implemented in THELI. A VISIR image processed with THELI is shown in 
Fig.~\ref{ngc1068}.

\section{\label{weighting}Weighting and defect detection}
The general weighting process has been described in detail in E05. To summarize, a weight map is created based on
the normalized flat. This \textit{global weight} is then modified individually to mask cosmics, hot pixels or 
satellites present in the images.

Many CCDs suffer from bad columns, some of which cannot be corrected using dark exposures. THELI 
is able to identify such defects (even weak ones) in a well exposed flat field. The variations in a 
flat field due to uneven illumination are often larger than those caused by bad columns. These effects 
must therefore first be corrected for, which is done by dividing the flat field by itself after convolution 
with a 30 pixel wide Gaussian. The mean value of an individual column is then compared to the overall 
mean of the image. The column will be masked if it deviates by more than a user-defined threshold 
(default: 2\%).

\section{Photometry and astrometry}
\subsection{\label{directphot}Direct photometric calibration}
\textit{Direct} photometric calibration means that the zero-point of an individual exposure is obtained 
from stars of known magnitude within the same field. Currently, SDSS-DR9 \citep{aaa12} and 2MASS \citep{scs06} 
catalogs are available in THELI for this purpose. A (configurable) maximum uncertainty of 0.05 mag is allowed 
for the reference magnitudes. Future deeper/wider surveys will be implemented as they become 
publicly available.

This method has three advantages compared to the classic \textit{indirect} calibration.
First, the data can be taken in arbitrary photometric conditions. Second, multi-chip cameras can be 
corrected for residual zero-point differences between chips \citep[see e.g.][]{shk11}. Third, no 
observing time has to be reserved for standard stars at different airmass. On the downside, 
possible color terms remain undetermined.

\subsection{Indirect photometric calibration}
In E05 we describe photometric calibration using standard stars and a three 
parameter fit (zero-point, color term, extinction). Only the bright $UBVRI$ standards by 
\cite{lan92} and \cite{ste00} were implemented at the time. Today THELI provides the SDSS stripe 82 
photometric calibration catalog \citep{ism07} in addition, covering 20:30:00 $<$ RA $<$ 04:00:00, and 
$-$01:16:00 $<$ DEC $<$ 01:16:00. Originally this catalog contains 1.01 million point sources, but
only objects with photometric errors $<0.05$ mag in all bands are kept. Very faint sources in the 
$gri$ filters, which could lead to a systematic bias of 0.02 mag in the zero-point, have also been removed.
$340\;000$ sources remain, all of which have $ugriz$ magnitudes in the SDSS 2.5m system as well as 
$u^\prime g^\prime r^\prime i^\prime z^\prime$ in the USNO 1.0m system \citep{sta07}. The catalog has to
be downloaded separately from the THELI webpage.

For near-IR data, $JHK_s$ magnitudes from \cite{lcv06} (UKIRT MKO), \cite{hmt98} and
\cite{pmk98} are available, as well as the 2MASS calibration fields by \cite{nws00}. A combined 
$YJHK_sLM$ catalog from UKIRT/JAC has also been added, together with $L^\prime M^\prime$ data from 
\cite{lhc03}.

\subsection{\label{astrometry}Astrometry}
Astronomical instruments cover very small to very large fields of view, hence absolute astrometric 
calibrations rely on suitably chosen reference catalogs. Extreme cases are e.g.~near-IR 
data of a small field with low source density, or wide-field images of the crowded galactic plane. 
Successful matching of source and reference catalogs requires sufficient mutual overlap. A variety of 
adjustments are available if matching with default values fails:

\begin{itemize}
\item{Depth of the reference catalog (to avoid computational limits for crowded fields)}
\item{Depth of the {\tt SExtractor} \citep{bea96} source catalog}
\item{Deblending of composite sources (e.g.~richly structured galaxies)}
\item{Filters to reject spurious sources}
\item{Various reference catalogs (e.g.~for greater depth or better wavelength match)}
\item{Different matching algorithms}
\end{itemize}

Currently implemented catalogs are SDSS-DR9, PPMXL, USNO-B1, 2MASS, UCAC-4, GSC-2.3 and TYCHO 
\citep[][respectively]{aaa12,rds10,mlc03,csd03,zfg12,lll08,esa97}. TYCHO can be used for images
taken with wide angle photo lenses covering hundreds of square degrees. For all-sky camera data,
a locally stored, filtered version of GSC-2.3 with an upper magnitude limit of 10 is available.

Sometimes these catalogs are insufficient, e.g.~if the target is a bright and richly 
structured nearby galaxy. The reference catalog might then be empty because the photographic 
plate was saturated, or the source deblending insufficient. In this case, secondary reference 
catalogs can be created with THELI from images with valid WCS headers taken with another 
telescope (see e.g.~Sect.~\ref{astrometrychallanges}).

The astrometric solution is stored in separate FITS headers, which are read by {\tt SWarp} 
\citep{ber10_2} during image coaddition. The individual images remain uncorrected, unless the 
\textit{Update header} function is used, inserting the first order solution ({\tt CRPIX1/2}, 
{\tt CRVAL1/2}, and {\tt CDij}) into the headers. This is optional and can be undone anytime. If 
fully distortion-corrected individual images are needed the resampled data (Sect.~\ref{outliercoadd}) 
can be utilized.

\subsubsection{\label{scamp}Scamp}
The most commonly used astrometric tool in THELI is {\tt Scamp} \citep{ber06}, developed in particular for 
multi-chip cameras. Previously stored information about the detectors' relative orientation (the 
\textit{focal plane}, hereafter FP) can be used. To this end the FP has has been measured for all pre-configured 
multi-chip cameras in THELI (Tables \ref{opticalinsts} and \ref{irinsts}, App.~\ref{insttables}) based on 
dense stellar fields. 
FP models may need updates from time to time, e.g.~if an instrument has undergone mechanical or optical 
maintenance. In cases where the detectors are not located in the same physical plane (e.g.~VIMOS@VLT, 
SPARTAN@SOAR), differential flexure between the optical arms may require the FP model to be updated
for each data set. This functionality is readily available in THELI.

The internal accuracy of the resulting astrometric solution with {\tt Scamp} is on the order of $0.06-0.12$ 
pixels for optical wide-field imagers \citep[see e.g.][]{ehm13}. For low density fields observed in the 
near-IR it may be reduced to $0.1-0.3$ pixels. Once the astrometric solution is obtained, {\tt Scamp} also 
determines relative photometric zero-points for the exposures.

{\tt Scamp} matches the object catalogs to the astrometric reference catalog as follows. In the first step,
the pixel scale and relative position angle are determined for both catalogs by 
\textit{``cross-correlating the 2D histograms 
of source pair coordinates in the log(separation) vs. position-angle space''} 
\citep[][]{ber06,kwl99}. The offset in RA and DEC is found by correlating the 2D positional histograms. 
This approach works well as long as the positional uncertainty is not too large, i.e.~the offset between 
nominal and true coordinates is not larger than about half the field of view. However, this condition is not 
always met, which can make the astrometric solution difficult (in particular if the position angle and a 
possible flip are still undetermined, e.g.~for commissioning data).

\subsubsection{Astrometry.net}
A complementary solution to {\tt Scamp} is a local implementation of the {\tt astrometry.net} algorithm
\citep{lhm10}. Matching is based on quadrilaterals in the reference and the object catalogs. {\tt astrometry.net} 
can create its own source catalogs, but in THELI it will use the same {\tt SExtractor} catalogs 
as created for {\tt Scamp}. In this way the user has more control over source densities and spurious 
detections. THELI does not use the {\tt astrometry.net} all-sky online index, but builds the (much smaller) 
reference index from the reference catalog. By running the {\tt astrometry.net} client locally, an exposure 
with a few hundred to a thousand sources is typically solved within a fraction of a second. 

The different matching techniques used by {\tt Scamp} and {\tt astrometry.net} complement each other 
for problematic fields. The disadvantages of the current implementation of {\tt astrometry.net} (v0.43)  
include its inability to calculate relative photometric zero-points, while also offering little control 
over how e.g.~distortion or chip alignment are handled, in particular for multi-chip cameras. {\tt astrometry.net} 
calculates the distortion polynomial coefficients in the SIP convention \citep{sls12}, which is not understood 
by {\tt SWarp} which uses the older PV convention. Hence {\tt astrometry.net} is currently used for catalog 
matching only, while {\tt Scamp} is run automatically afterwards (with matching deactivated) to calculate the 
distortion maps and relative photometric zero-points.

\subsubsection{Shift and cross-correlation for mid-IR data}
Both {\tt Scamp} and {\tt astrometry.net} deliver full WCS solutions. However, they need sufficiently 
high object density to constrain all parameters. This is never the case for mid-IR data, where fields of 
view are (1) on the order of a few tens of arcseconds, and (2) contain only one or a few sources (possibly all 
non-stellar). In addition, suitable all-sky mid-IR catalogs with sufficient spatial resolution are not 
available. Thus for mid-IR data, THELI simply measures the linear shift between exposures in image 
coordinates, and constructs a dummy WCS solution that is understood by {\tt SWarp} for stacking. 

Two methods are available. In the first, the \textit{Shift} approach is based on {\tt SExtractor} 
catalogs and recommended for data in which one (or a few) point source(s) can be repeatedly
detected in all images. Relative photometric zero-points are determined as well. In the second method,
the offsets are determined using 2D cross-correlation of noise clipped images. This approach is recommended 
for data with predominantly extended flux. Relative photometry is \textit{not} obtained, which is of little 
concern as mid-IR observations require clear and dry conditions. The WCS of the 
coadded images still has to be corrected manually, as global astrometry is not performed.

\subsubsection{Adopting the original WCS header}
If an astrometric solution cannot be found with either of the above methods, then the user can simply 
adopt the zero-order WCS solution already present in the raw FITS headers. Only {\tt CRPIX1/2}, 
{\tt CRVAL1/2} and the {\tt CD}-matrix will be copied in this case, while distortion terms are ignored 
and relative photometric zero-points not calculated. This may be useful if the relative dither offsets 
between exposures are known to be precisely reflected in the headers.

\begin{figure*}[t]
  \includegraphics[width=1.0\hsize]{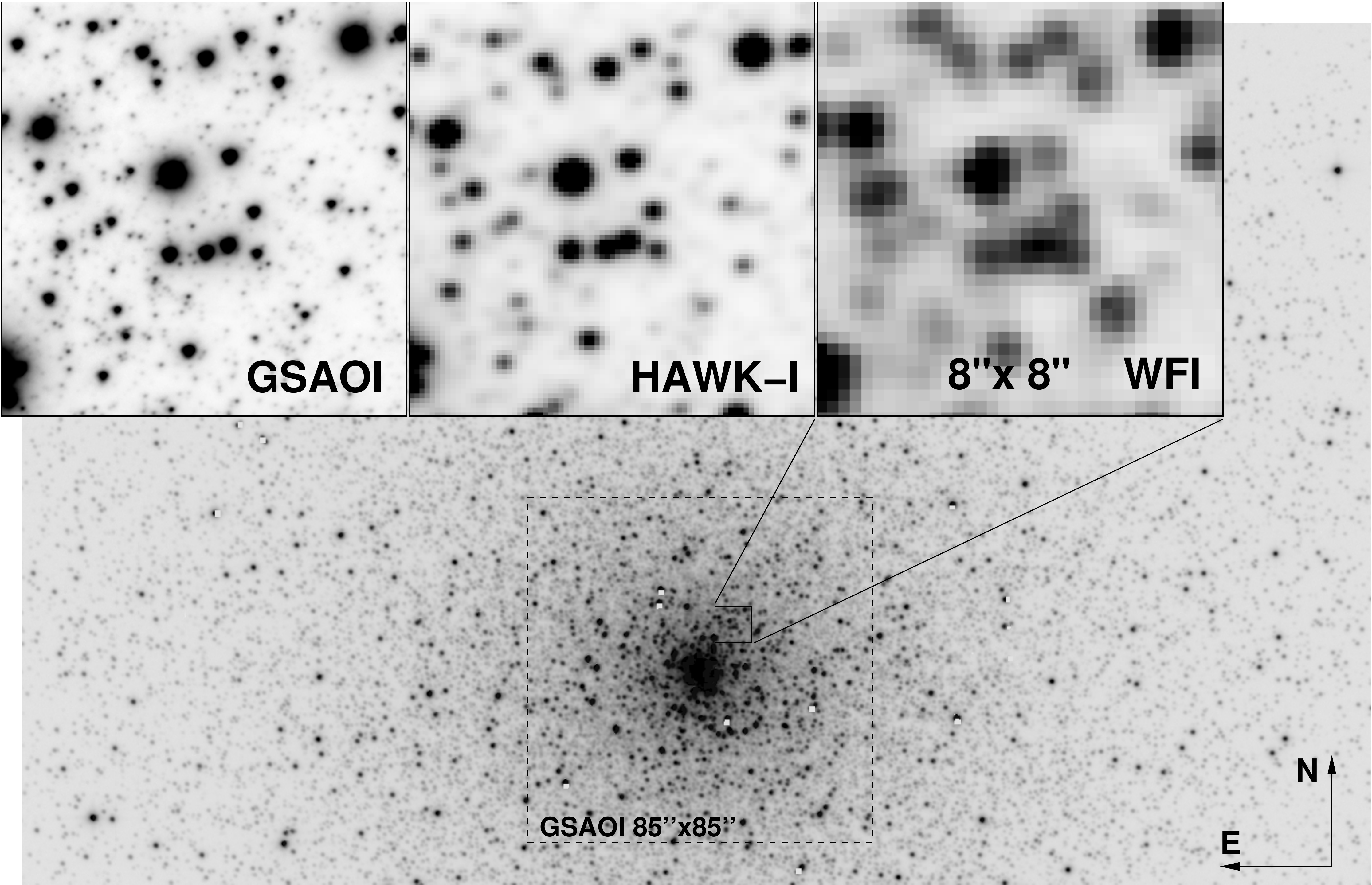}
  \caption{\label{ngc1851}{MCAO data pose a new challenge for astrometric calibration. Given the insufficient
      resolution of available reference catalogs, secondary or even tertiary catalogs are necessary (see text 
      for details). The WFI/HAWK-I/GSAOI data in this example have a FWHM of 0\myarcsec75, 0\myarcsec33 and 
      0\myarcsec08, and pixel scales of 0\myarcsec24, 0\myarcsec10 and 0\myarcsec02, respectively. All data 
      were reduced with THELI.}}
\end{figure*}

\subsubsection{\label{astrometrychallanges}An astrometric challenge: AO images}
Current astrometric reference catalogs (e.g.~USNO, 2MASS) are mostly based on plate scans or wide field 
images with an angular resolution on the order of $1^{\prime\prime}-2^{\prime\prime}$. This is
about $3-5$ times lower than the resolution of typical seeing-limited images, and does not pose a 
problem as long as fields are not very crowded. However, the limit is reached with adaptive optics (AO) 
systems such as GeMS/GSAOI@Gemini \citep{mhs04}, and similar instrumentation at future extremely large 
telescopes. The pixel scale of GSAOI is 0\myarcsec0197, 50 times smaller than the resolution element of the 
Digitized Sky Survey. Hence the source density of the plate scans becomes too low for successful catalog 
matching. Resolved multiple sources in the AO data introduce ambiguities, complicating the matching 
process further. The fields of view of current multi-conjugated AO (MCAO) systems are on the order of 
1$^\prime$, and thus of the same size or smaller than many of the prominent science targets that will 
be revisited with these instruments. A global astrometric calibration of such data requires high 
resolution (and possibly much deeper) secondary catalogs obtained from images taken with 
2m$-$8m class telescopes.

Figure \ref{ngc1851} of globular cluster NGC 1851 illustrates the aforementioned challenges. The high source 
density in this field required a staggered approach. For the successful astrometry of the GSAOI data a tertiary 
catalog had to be used, obtained from a high resolution $K_s$-band HAWK-I@VLT image. This, in turn, could not be 
calibrated with USNO-B1 (no detections within the cluster) or 2MASS (too shallow in the outskirts, too few 
detections within the cluster). Instead, a reference catalog was extracted from an optical wide field image 
taken with WFI at the 2.2m MPG/ESO telescope, which itself was matched against 2MASS. At the other extreme we 
find observations of extragalactic targets, where hardly any reference sources are available in the common 
catalogs within a 1$^\prime$ radius. Deep classical observations in good seeing are thus required to provide 
a secondary standard star catalog.

\section{\label{sky}Sky subtraction}
The main objective of sky subtraction is to achieve a homogeneous zero background level across the FP.
At this point THELI assumes that any background variations are additive and hence must be subtracted, 
i.e.~that the photometric zero-point is already uniform.

Sky subtraction is essential for multi-chip cameras. Consider two different pixels in a coadded image,
constructed from $n$ dithered exposures without sky subtraction. While all $n$ exposures contribute 
to the first coadded pixel, the second pixel may lack data points due to detector gaps. If the background 
level varies between exposures, then their mean (or median) values at the position of the two coadded pixels
will be different. Consequently, chip gaps will show up as discrete brighter or darker areas in the coadded 
image without prior sky subtraction.

Difficulties arise for crowded fields or extended objects (with faint halos), where local and unbiased 
background measurements are difficult. Gradients caused by e.g.~zodiacal light or the moon cause further
complications. These can be controlled earlier on with separate sky exposures (see Sect.~\ref{background}), 
and/or with the the following two options for individual sky subtraction.

\subsection{\label{skymodel}Full modeling}
For sparse fields and fields without extended objects full background modeling can be done. The 
process is similar to what is described in E05, with some modifications. {\tt SExtractor} is used to 
mask objects, with adjustable detection thresholds. To some degree a faint extended halo can be masked, as 
long as the amplitude of the background variations is smaller than that of the halo. The masked areas are 
interpolated iteratively, based on the values of non-masked pixels in the local neighborhood. In this way 
background variations are reflected properly across larger masks, where a single constant estimate is 
insufficient. The resulting image, free of objects, is then convolved with a Gaussian kernel 
yielding the final sky model. Ideally, the kernel's FWHM is equal to or smaller than the typical extent 
of the background variations, yet large enough so that the filter remains insensitive to extended 
sources.

\subsection{Subtracting a constant value}
Background modeling on individual images is not an option whenever surface photometry or detection
of faint extended structures is required. This holds in particular if the structures are not visible 
in individual images. Intra-cluster light, tidal features, and faint halos around brighter objects are 
good examples \citep{rzm08,tsl09,mgc10,gar12}. To preserve the faint structures a constant sky must be 
subtracted.

In THELI a constant background value can be determined individually for each chip of a detector mosaic,
using various statistical estimators. Alternatively, a particular chip unaffected by the extended object can be 
used to estimate the sky level for all other chips. The measurement is performed either from the
entire chip or from a sub-area, and after object masking has taken place (Sect.~\ref{skymodel}).
Sub-areas can be defined using pixel coordinates (and are thus fixed to the detector), or by
sky coordinates such that the measurement box moves with the dither pattern. The latter is useful 
for crowded fields, allowing the background to be measured repeatedly and precisely from the same 
location. Note that this (alternative) requires the \textit{Update header} function to have been applied
after the successful calculation of an astrometric solution (Sect.~\ref{astrometry}).

Should even more flexibility be required, the constant background values determined can be adjusted 
manually for individual chips or entire mosaics until the coaddition is satisfactory.

\section{Coaddition}
During the image coaddition phase the user can choose between the following {\tt SWarp} parameters:

\begin{itemize}
\item{Celestial coordinate systems (equatorial, galactic, ecliptic, supergalactic)}
\item{WCS projections (TAN, COE, ...)}
\item{Combine types (weighted mean, median, ...)}
\item{Resampling kernels}
\item{Output pixel scale}
\end{itemize}

For multi-color data sets it is recommended to use identical reference coordinates (RA, DEC) for 
the sky projection. A particular object will then end up on precisely the same pixel in the coadded 
images of all different filters (see App.~\ref{prepcolor}). For non-equatorial coordinate systems the 
corresponding counterparts must be used, e.g.~galactic longitude and latitude.

\subsection{Filtering and arbitrary sky position angles}
In the first step, {\tt SWarp} gets a list of all images to stack, and creates a FITS header for the 
coadded image. At this level the user can decide which exposures enter the coaddition process, be it a
selection by filter (if a multi-color data set is present), seeing, and/or relative photometric zero-point.
For multi-chip cameras a selection of chips can also be made, resulting in only part of the mosaic being 
created. 

Two additional options act directly on the header of the coadded image before it is created. In case of 
wide-angle and all-sky projections it is difficult to predict the geometry of the coadded image, resulting 
in truncation. In this case the output size ({\tt NAXIS1/2}) can be corrected. The other option is 
to choose an arbitrary sky position angle, as {\tt SWarp} by default orients all stacked images with north 
up and east to the left. This is done by updating the {\tt CD}-matrix, and increasing the image geometry 
to accommodate the new layout.

\subsubsection{\label{outliercoadd}Outlier rejection}
In the second step, {\tt SWarp} resamples the individual images, normally followed by the coaddition.
{\tt SWarp} does not however offer outlier rejection during stacking. While a median filter may 
be used to reject outliers during coaddition, its variance is a factor of $\sim\pi/2$ higher compared
to a mean combination \citep[assuming Gaussian noise in the input images, see Sect. 3.13 of][]{kai02}. 
The effective exposure time would need to be increased by a similar factor to compensate for this 
loss of sensitivity. THELI therefore offers the option of reconstructing
the stacks for each coadded pixel based on the resampled images, identifying bad pixels by running a 
$\sigma$-clipping algorithm. Spurious pixels have then their weight set to zero in the resampled weight 
images. In this way photometric integrity is preserved while still using the weighted combination.

If a pixel is found to be bad, it may be masked only if the $k$ neighboring pixels are also bad (\textit{bad 
pixel cluster size}). This accounts for the fact that most resampling kernels distribute the flux of 
one input pixel over several output pixels. In addition, if a pixel is located $m$ pixels or less from a 
bad pixel, it can also be masked. This is useful for the complete masking of bright asteroids or satellites, 
whose faint wings caused by the PSF go undetected by the outlier rejection.

\subsubsection{Locking onto proper motion targets}
\begin{figure}[t]
  \includegraphics[width=1.0\hsize]{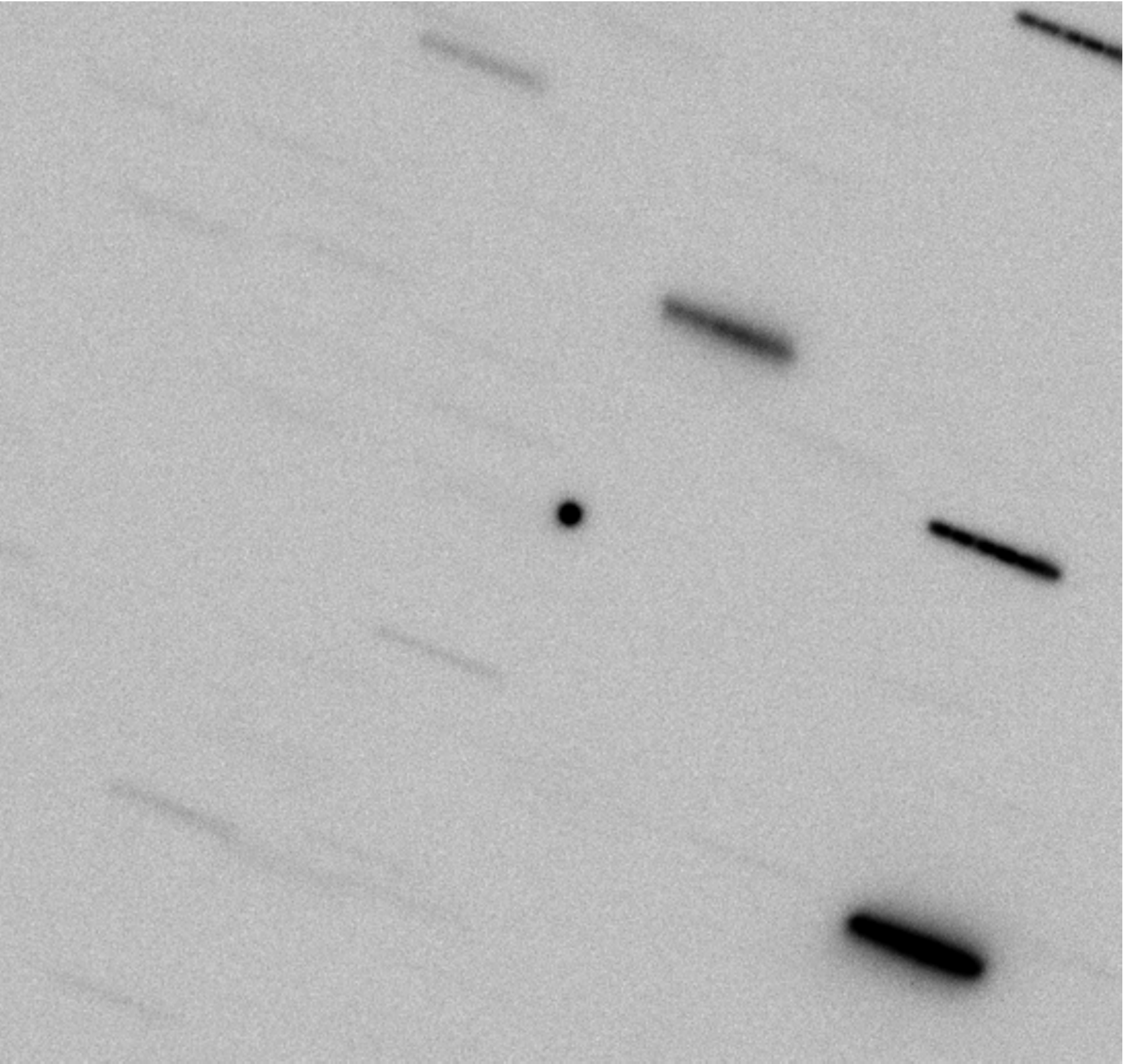}
  \caption{\label{tno}{TNO 1996 TO$_{66}$, as observed with FORS1@VLT using sidereal tracking (data from 
      \cite{sbh02}). The proper motion vector was applied during image coaddition.}}
\end{figure}

The proper motion vector of solar system objects can be projected into the headers of the resampled images, 
before final coaddition, by adjusting the {\tt CRVAL1/2} keywords. The coadded image then follows the 
moving target, whereas stars appear trailed. In this manner faint moving objects, possibly invisible 
in single images, can be analyzed, provided their proper motions are known or can be guessed. The 
{\tt DATE-OBS} keyword must be present in the input headers for sufficiently accurate timing. Currently, 
only linear proper motion vectors are supported, while second order effects are ignored. Figure \ref{tno} 
shows an example image.

\subsubsection{Edge smoothing}
In some cases it is not possible to achieve good sky subtraction, resulting in small discontinuous jumps at 
chip boundaries. This can be problematic for surface photometry or feature detection, in particular if the 
jump runs across the target. The discontinuity can be suppressed by \textit{edge smoothing}, where the edges 
of the weight images are softened with a sine function of wavelength $\lambda$ prior to resampling. Pixels 
at the outermost edge of the weight map are assigned a value of zero, corresponding to the minimum of the 
sine function. The pixel values then gradually increase inwards until they reach their original value after 
$0.5\lambda$ (maximum of the sine function). Within this border the weight maps are unchanged. The 2nd 
derivatives at the transition points are zero. The chosen \textit{edge smoothing length}, $0.5\lambda$, 
should be similar or a bit smaller than the size of the dither pattern. Photometry is conserved since this
method is applied to the weight maps only. The noise level along the chip boundaries is increased
by this process due to the lower weight.

\section{Summary and outlook}
In this paper I have presented an overview of the THELI data reduction pipeline. THELI integrates a large 
range of tools, and offers them in a homogeneous and convenient manner for end-to-end processing. Many different 
instruments at various observatories are pre-configured, and their data easily reduced. A short learning curve 
ensures quick success for new and more seasoned observers alike. The great flexibility facilitates a broad 
range of scientific studies, as demonstrated by e.g. \cite{maa05,scm08,llh12,mmb12,gjs13} and
\cite{lsr13}.

The development of THELI continues, as both astronomical instrumentation and software are highly 
dynamic areas. One of the main work areas, as already mentioned before, is a port to the newer {\tt Qt5} library 
ensuring cross-platform compatibility and the latest standard in GUI development. This next major release
will also include automatic satellite detection and a method for 2D illumination correction (see Sect. 
\ref{background}). On the instrumental side, the focus is on mid-IR cameras and further support for 
instruments and observatories that are not yet included.

\begin{acknowledgements}
\section*{Acknowledgements}
I thank Douglas Applegate, Abdul Azim, Rafael Barrena, Yuri Beletsky, Albert Bosma, Andrew 
Card\-well, Cristiano daRocha, Matthias Frank, Javier Fuentes, Chon Gayoung, Karianne Holhjem, Richard Hook, 
Erik Kool, Andrea Kunder, Sean Lake, 
Richard Lane, Pierre Leisy, Stefan Lieder, Thorsten Lisker, Ed Loh, Grigoris Maravelias, Davide de Martin, 
James McCormac, Steffen Mieske, Oleg Maliy, Carsten Moos, Richard M\"uller, Felipe Navarete, Nadine Neumeyer, 
Daniela Olave Rojas, Emanuela Pompei, David Russell, Julia Scharw\"achter, Oliver Sch\"utz, Zeinab Shafiee, 
Zahra Sheikhbahaee, Tim Schrabback, Patr\'{i}cia Figueir\'{o} Spinelli, Milan Stojanovic, Ignacio Toledo, 
Thomas Tuchan, 
Alex Tudorica, Daniela Wuttke, and Umut Yildiz for their numerous suggestions, testing, discussions, and access 
to data of new instruments. Their contributions resulted in the current functionality, support of many different 
instruments, and user friendliness of THELI. I apologize to all other people who contributed their ideas and 
feature requests since 2004 and whose names I forgot. Special thanks goes to Thomas Erben for 
maintaining the THELI pipeline core, and to Emmanuel Bertin for his continuous development and support of 
{\tt SExtractor}, {\tt SWarp} and {\tt Scamp}. Lastly, I would like to thank the anonymous referee for his 
very helpful comments about the manuscript, and Karianne Holhjem for substantial language editing.
\end{acknowledgements}

\bibliography{mybib}

\begin{thebibliography}{69}
\expandafter\ifx\csname natexlab\endcsname\relax\def\natexlab#1{#1}\fi

\bibitem[{{Ahn} {et~al.}(2012){Ahn}, {Alexandroff}, {Allende Prieto},
  {Anderson}, {Anderton}, {Andrews}, {Aubourg}, {Bailey}, {Balbinot}, {Barnes},
  \& et~al.}]{aaa12}
{Ahn}, C.~P., {Alexandroff}, R., {Allende Prieto}, C., {et~al.} 2012, \apjs,
  203, 21

\bibitem[{{Albert} {et~al.}(2008){Albert}, {Baril}, {Ward}, {Arnouts}, \&
  {Devost}}]{abw08}
{Albert}, L., {Baril}, M., {Ward}, J., {Arnouts}, S., \& {Devost}, D. 2008,
  Proc. SPIE, 7014, 85

\bibitem[{{Baade} {et~al.}(1999){Baade}, {Meisenheimer}, {Iwert}, {Alonso},
  {Augusteijn}, {Beletic}, {Bellemann}, {Benesch}, {B{\"o}hm}, {B{\"o}hnhardt},
  {Brewer}, {Deiries}, {Delabre}, {Donaldson}, {Dupuy}, {Franke}, {Gerdes},
  {Gilliotte}, {Grimm}, {Haddad}, {Hess}, {Ihle}, {Klein}, {Lenzen}, {Lizon},
  {Mancini}, {M{\"u}nch}, {Pizarro}, {Prado}, {Rahmer}, {Reyes}, {Richardson},
  {Robledo}, {Sanchez}, {Silber}, {Sinclaire}, {Wackermann}, \&
  {Zaggia}}]{bmi99}
{Baade}, D., {Meisenheimer}, K., {Iwert}, O., {et~al.} 1999, The Messenger, 95,
  15

\bibitem[{{Banse} {et~al.}(1983){Banse}, {Crane}, {Grosbol}, {Middleburg},
  {Ounnas}, {Ponz}, \& {Waldthausen}}]{bcg83}
{Banse}, K., {Crane}, P., {Grosbol}, P., {et~al.} 1983, The Messenger, 31, 26

\bibitem[{{Beckett} {et~al.}(1997){Beckett}, {Mackay}, {McMahon}, {Parry},
  {Piche}, \& {Ellis}}]{bmm97}
{Beckett}, M.~G., {Mackay}, C.~D., {McMahon}, R.~G., {et~al.} 1997, Proc. SPIE,
  2871, 1152

\bibitem[{{Bertin}(2006)}]{ber06}
{Bertin}, E. 2006, in ASP Conf. Ser., Vol. 351, ADASS XV, ed. C.~{Gabriel},
  C.~{Arviset}, D.~{Ponz}, \& S.~{Enrique}, 112

\bibitem[{Bertin(2010)}]{ber10_2}
Bertin, E. 2010, Astrophysics Source Code Library, 10068

\bibitem[{Bertin \& Arnouts(1996)}]{bea96}
Bertin, E. \& Arnouts, S. 1996, A\&AS, 117, 393

\bibitem[{{Butcher} \& {Stevens}(1981)}]{bus81}
{Butcher}, H. \& {Stevens}, R. 1981, Kitt Peak National Observatory Newsletter,
  16, 6

\bibitem[{{Crane} {et~al.}(1981){Crane}, {Rose}, \& {Schabel}}]{crs81}
{Crane}, P., {Rose}, A., \& {Schabel}, P. 1981, Proc. SPIE, 290, 120

\bibitem[{{Cuillandre} {et~al.}(2000){Cuillandre}, {Luppino}, {Starr}, \&
  {Isani}}]{cls00}
{Cuillandre}, J.-C., {Luppino}, G.~A., {Starr}, B.~M., \& {Isani}, S. 2000,
  Proc. SPIE, 4008, 1010

\bibitem[{{Cutri} {et~al.}(2003){Cutri}, {Skrutskie}, {van Dyk}, {Beichman},
  {Carpenter}, {Chester}, {Cambresy}, {Evans}, {Fowler}, {Gizis}, {Howard},
  {Huchra}, {Jarrett}, {Kopan}, {Kirkpatrick}, {Light}, {Marsh}, {McCallon},
  {Schneider}, {Stiening}, {Sykes}, {Weinberg}, {Wheaton}, {Wheelock}, \&
  {Zacarias}}]{csd03}
{Cutri}, R.~M., {Skrutskie}, M.~F., {van Dyk}, S., {et~al.} 2003, {2MASS All
  Sky Catalog of point sources.} (CDS)

\bibitem[{{Da Rocha} {et~al.}(2008){Da Rocha}, {Ziegler}, \& {Mendes de
  Oliveira}}]{rzm08}
{Da Rocha}, C., {Ziegler}, B.~L., \& {Mendes de Oliveira}, C. 2008, \mnras,
  388, 1433

\bibitem[{{Erben} {et~al.}(2013){Erben}, {Hildebrandt}, {Miller}, {van
  Waerbeke}, {Heymans}, {Hoekstra}, {Kitching}, {Mellier}, {Benjamin}, {Blake},
  {Bonnett}, {Cordes}, {Coupon}, {Fu}, {Gavazzi}, {Gillis}, {Grocutt}, {Gwyn},
  {Holhjem}, {Hudson}, {Kilbinger}, {Kuijken}, {Milkeraitis}, {Rowe},
  {Schrabback}, {Semboloni}, {Simon}, {Smit}, {Toader}, {Vafaei}, {van Uitert},
  \& {Velander}}]{ehm13}
{Erben}, T., {Hildebrandt}, H., {Miller}, L., {et~al.} 2013, \mnras, 433, 2545

\bibitem[{Erben {et~al.}(2005)Erben, Schirmer, Dietrich, {et~al.}}]{esd05}
Erben, T., Schirmer, M., Dietrich, J., {et~al.} 2005, AN, 326, 432

\bibitem[{{Flaugher} {et~al.}(2012){Flaugher}, {Abbott}, {Angstadt}, {Annis},
  {Antonik}, {Bailey}, {Ballester}, {Bernstein}, {Bernstein}, {Bonati},
  {Bremer}, {Briones}, {Brooks}, {Buckley-Geer}, {Campa}, {Cardiel-Sas},
  {Castander}, {Castilla}, {Cease}, {Chappa}, {Chi}, {da Costa}, {DePoy},
  {Derylo}, {de Vincente}, {Diehl}, {Doel}, {Estrada}, {Eiting}, {Elliott},
  {Finley}, {Flores}, {Frieman}, {Gaztanaga}, {Gerdes}, {Gladders}, {Guarino},
  {Gutierrez}, {Grudzinski}, {Hanlon}, {Hao}, {Holland}, {Honscheid},
  {Huffman}, {Jackson}, {Jonas}, {Karliner}, {Kau}, {Kent}, {Kozlovsky},
  {Krempetz}, {Krider}, {Kubik}, {Kuehn}, {Kuhlmann}, {Kuk}, {Lahav},
  {Langellier}, {Lathrop}, {Lewis}, {Lin}, {Lorenzon}, {Martinez}, {McKay},
  {Merritt}, {Meyer}, {Miquel}, {Morgan}, {Moore}, {Moore}, {Neilsen}, {Nord},
  {Ogando}, {Olson}, {Patton}, {Peoples}, {Plazas}, {Qian}, {Roe}, {Roodman},
  {Rossetto}, {Sanchez}, {Soares-Santos}, {Scarpine}, {Schalk}, {Schindler},
  {Schmidt}, {Schmitt}, {Schubnell}, {Schultz}, {Selen}, {Serrano}, {Shaw},
  {Simaitis}, {Slaughter}, {Smith}, {Spinka}, {Stefanik}, {Stuermer},
  {Sypniewski}, {Talaga}, {Tarle}, {Thaler}, {Tucker}, {Walker}, {Weaverdyck},
  {Wester}, {Woods}, {Worswick}, \& {Zhao}}]{faa12}
{Flaugher}, B.~L., {Abbott}, T.~M.~C., {Angstadt}, R., {et~al.} 2012, Proc.
  SPIE, 8446, 11

\bibitem[{{Freyhammer} {et~al.}(2001){Freyhammer}, {Andersen}, {Arentoft},
  {Sterken}, \& {N{\o}rregaard}}]{faa01}
{Freyhammer}, L.~M., {Andersen}, M.~I., {Arentoft}, T., {Sterken}, C., \&
  {N{\o}rregaard}, P. 2001, Experimental Astronomy, 12, 147

\bibitem[{{Gentile} {et~al.}(2013){Gentile}, {J{\'o}zsa}, {Serra}, {Heald}, {de
  Blok}, {Fraternali}, {Patterson}, {Walterbos}, \& {Oosterloo}}]{gjs13}
{Gentile}, G., {J{\'o}zsa}, G.~I.~G., {Serra}, P., {et~al.} 2013, \aap, 554,
  A125

\bibitem[{{Goad} \& {Ball}(1981)}]{gob81}
{Goad}, L.~E. \& {Ball}, W.~F. 1981, Proc. SPIE, 290, 130

\bibitem[{{Guennou} {et~al.}(2012){Guennou}, {Adami}, {Da Rocha}, {Durret},
  {Ulmer}, {Allam}, {Basa}, {Benoist}, {Biviano}, {Clowe}, {Gavazzi},
  {Halliday}, {Ilbert}, {Johnston}, {Just}, {Kron}, {Kubo}, {Le Brun},
  {Marshall}, {Mazure}, {Murphy}, {Pereira}, {Raba{\c c}a}, {Rostagni},
  {Rudnick}, {Russeil}, {Schrabback}, {Slezak}, {Tucker}, \&
  {Zaritsky}}]{gar12}
{Guennou}, L., {Adami}, C., {Da Rocha}, C., {et~al.} 2012, \aap, 537, A64

\bibitem[{{Gunn} \& {Westphal}(1981)}]{guw81}
{Gunn}, J.~E. \& {Westphal}, J.~A. 1981, Proc. SPIE, 290, 16

\bibitem[{{Heymans} {et~al.}(2012){Heymans}, {Van Waerbeke}, {Miller}, {Erben},
  {Hildebrandt}, {Hoekstra}, {Kitching}, {Mellier}, {Simon}, {Bonnett},
  {Coupon}, {Fu}, {Harnois D{\'e}raps}, {Hudson}, {Kilbinger}, {Kuijken},
  {Rowe}, {Schrabback}, {Semboloni}, {van Uitert}, {Vafaei}, \&
  {Velander}}]{hwm12}
{Heymans}, C., {Van Waerbeke}, L., {Miller}, L., {et~al.} 2012, \mnras, 427,
  146

\bibitem[{Hildebrandt {et~al.}(2006)Hildebrandt, Erben, Dietrich, Cordes,
  Haberzettl, Hetterscheidt, Schirmer, Schmithuesen, Schneider, Simon, \&
  Trachetrnach}]{hed06}
Hildebrandt, H., Erben, T., Dietrich, J.~P., {et~al.} 2006, A\&A, 452, 1121

\bibitem[{{Hummel} {et~al.}(2010){Hummel}, {Hanuschik}, {de Bilbao}, {Mieske},
  {Szeifert}, {Ivanov}, \& {Castro}}]{hhb10}
{Hummel}, W., {Hanuschik}, R., {de Bilbao}, L., {et~al.} 2010, Proc. SPIE,
  7737, 35

\bibitem[{{Hunt} {et~al.}(1998){Hunt}, {Mannucci}, {Testi}, {Migliorini},
  {Stanga}, {Baffa}, {Lisi}, \& {Vanzi}}]{hmt98}
{Hunt}, L.~K., {Mannucci}, F., {Testi}, L., {et~al.} 1998, \aj, 115, 2594

\bibitem[{{Ivezi{\'c}} {et~al.}(2007){Ivezi{\'c}}, {Smith}, {Miknaitis}, {Lin},
  {Tucker}, {Lupton}, {Gunn}, {Knapp}, {Strauss}, {Sesar}, {Doi}, {Tanaka},
  {Fukugita}, {Holtzman}, {Kent}, {Yanny}, {Schlegel}, {Finkbeiner},
  {Padmanabhan}, {Rockosi}, {Juri{\'c}}, {Bond}, {Lee}, {Stoughton}, {Jester},
  {Harris}, {Harding}, {Morrison}, {Brinkmann}, {Schneider}, \& {York}}]{ism07}
{Ivezi{\'c}}, {\v Z}., {Smith}, J.~A., {Miknaitis}, G., {et~al.} 2007, \aj,
  134, 973

\bibitem[{{Jacoby} {et~al.}(2002){Jacoby}, {Tonry}, {Burke}, {Claver}, {Starr},
  {Saha}, {Luppino}, \& {Harmer}}]{jtb02}
{Jacoby}, G.~H., {Tonry}, J.~L., {Burke}, B.~E., {et~al.} 2002, Proc. SPIE,
  4836, 217

\bibitem[{Kaiser(2002)}]{kai02}
Kaiser, N. 2002, Elements of Astrophysics (University of Hawaii, available
  online at {\tt http://www.ifa.hawaii.edu/$\sim$kaiser/
  lectures/elements.pdf})

\bibitem[{Kaiser {et~al.}(1999)Kaiser, Wilson, Luppino, \& Dahle}]{kwl99}
Kaiser, N., Wilson, G., Luppino, G., \& Dahle, H. 1999, preprint
  astro-ph/9907229

\bibitem[{{Koch} {et~al.}(2004){Koch}, {Odenkirchen}, {Grebel}, \&
  {Caldwell}}]{kog04}
{Koch}, A., {Odenkirchen}, M., {Grebel}, E.~K., \& {Caldwell}, J.~A.~R. 2004,
  Astronomische Nachrichten, 325, 299

\bibitem[{{Kuijken} {et~al.}(2002){Kuijken}, {Bender}, {Cappellaro},
  {Muschielok}, {Baruffolo}, {Cascone}, {Iwert}, {Mitsch}, {Nicklas},
  {Valentijn}, {Baade}, {Begeman}, {Bortolussi}, {Boxhoorn}, {Christen},
  {Deul}, {Geimer}, {Greggio}, {Harke}, {H{\"a}fner}, {Hess}, {Hess}, {Hopp},
  {Ilijevski}, {Klink}, {Kravcar}, {Lizon}, {Magagna}, {M{\"u}ller},
  {Niemeczek}, {de Pizzol}, {Poschmann}, {Reif}, {Rengelink}, {Reyes},
  {Silber}, \& {Wellem}}]{kpc02}
{Kuijken}, K., {Bender}, R., {Cappellaro}, E., {et~al.} 2002, The Messenger,
  110, 15

\bibitem[{{Landolt}(1992)}]{lan92}
{Landolt}, A.~U. 1992, \aj, 104, 340

\bibitem[{{Lane} {et~al.}(2013){Lane}, {Salinas}, \& {Richtler}}]{lsr13}
{Lane}, R.~R., {Salinas}, R., \& {Richtler}, T. 2013, \aap, 549, A148

\bibitem[{{Lang} {et~al.}(2010){Lang}, {Hogg}, {Mierle}, {Blanton}, \&
  {Roweis}}]{lhm10}
{Lang}, D., {Hogg}, D.~W., {Mierle}, K., {Blanton}, M., \& {Roweis}, S. 2010,
  \aj, 139, 1782

\bibitem[{{Lasker} {et~al.}(2008){Lasker}, {Lattanzi}, {McLean}, {Bucciarelli},
  {Drimmel}, {Garcia}, {Greene}, {Guglielmetti}, {Hanley}, {Hawkins},
  {Laidler}, {Loomis}, {Meakes}, {Mignani}, {Morbidelli}, {Morrison},
  {Pannunzio}, {Rosenberg}, {Sarasso}, {Smart}, {Spagna}, {Sturch},
  {Volpicelli}, {White}, {Wolfe}, \& {Zacchei}}]{lll08}
{Lasker}, B.~M., {Lattanzi}, M.~G., {McLean}, B.~J., {et~al.} 2008, \aj, 136,
  735

\bibitem[{{Leggett} {et~al.}(2006){Leggett}, {Currie}, {Varricatt}, {Hawarden},
  {Adamson}, {Buckle}, {Carroll}, {Davies}, {Davis}, {Kerr}, {Kuhn}, {Seigar},
  \& {Wold}}]{lcv06}
{Leggett}, S.~K., {Currie}, M.~J., {Varricatt}, W.~P., {et~al.} 2006, \mnras,
  373, 781

\bibitem[{{Leggett} {et~al.}(2003){Leggett}, {Hawarden}, {Currie}, {Adamson},
  {Carroll}, {Kerr}, {Kuhn}, {Seigar}, {Varricatt}, \& {Wold}}]{lhc03}
{Leggett}, S.~K., {Hawarden}, T.~G., {Currie}, M.~J., {et~al.} 2003, \mnras,
  345, 144

\bibitem[{{Lieder} {et~al.}(2012){Lieder}, {Lisker}, {Hilker}, {Misgeld}, \&
  {Durrell}}]{llh12}
{Lieder}, S., {Lisker}, T., {Hilker}, M., {Misgeld}, I., \& {Durrell}, P. 2012,
  \aap, 538, A69

\bibitem[{Magnier \& Cuillandre(2004)}]{mac04}
Magnier, G. \& Cuillandre, J.-C. 2004, \pasp, 116, 449

\bibitem[{{Manfroid} {et~al.}(2001){Manfroid}, {Selman}, \& {Jones}}]{mas01}
{Manfroid}, J., {Selman}, F., \& {Jones}, H. 2001, The Messenger, 104, 16

\bibitem[{{Mart{\'{\i}}nez-Delgado} {et~al.}(2010){Mart{\'{\i}}nez-Delgado},
  {Gabany}, {Crawford}, {Zibetti}, {Majewski}, {Rix}, {Fliri},
  {Carballo-Bello}, {Bardalez-Gagliuffi}, {Pe{\~n}arrubia}, {Chonis}, {Madore},
  {Trujillo}, {Schirmer}, \& {McDavid}}]{mgc10}
{Mart{\'{\i}}nez-Delgado}, D., {Gabany}, R.~J., {Crawford}, K., {et~al.} 2010,
  \aj, 140, 962

\bibitem[{{McGregor} {et~al.}(2004){McGregor}, {Hart}, {Stevanovic}, {Bloxham},
  {Jones}, {Van Harmelen}, {Griesbach}, {Dawson}, {Young}, \& {Jarnyk}}]{mhs04}
{McGregor}, P., {Hart}, J., {Stevanovic}, D., {et~al.} 2004, Proc. SPIE, 5492,
  1033

\bibitem[{{McLean} {et~al.}(1981){McLean}, {Cormack}, {Herd}, \&
  {Aspin}}]{mch81}
{McLean}, I.~S., {Cormack}, W.~A., {Herd}, J.~T., \& {Aspin}, C. 1981, Proc.
  SPIE, 290, 155

\bibitem[{{McMichael} \& {Bentley}(2012)}]{mmb12}
{McMichael}, R.~T. \& {Bentley}, J. 2012, Proc. SPIE, 8486, 2

\bibitem[{{Meech} {et~al.}(2005){Meech}, {Ageorges}, {A'Hearn}, {Arpigny},
  {Ates}, {Aycock}, {Bagnulo}, {Bailey}, {Barber}, {Barrera}, {Barrena},
  {Bauer}, {Belton}, {Bensch}, {Bhattacharya}, {Biver}, {Blake},
  {Bockel{\'e}e-Morvan}, {Boehnhardt}, {Bonev}, {Bonev}, {Buie}, {Burton},
  {Butner}, {Cabanac}, {Campbell}, {Campins}, {Capria}, {Carroll}, {Chaffee},
  {Charnley}, {Cleis}, {Coates}, {Cochran}, {Colom}, {Conrad}, {Coulson},
  {Crovisier}, {deBuizer}, {Dekany}, {de L{\'e}on}, {Dello Russo}, {Delsanti},
  {DiSanti}, {Drummond}, {Dundon}, {Etzel}, {Farnham}, {Feldman},
  {Fern{\'a}ndez}, {Filipovic}, {Fisher}, {Fitzsimmons}, {Fong}, {Fugate},
  {Fujiwara}, {Fujiyoshi}, {Furusho}, {Fuse}, {Gibb}, {Groussin}, {Gulkis},
  {Gurwell}, {Hadamcik}, {Hainaut}, {Harker}, {Harrington}, {Harwit},
  {Hasegawa}, {Hergenrother}, {Hirst}, {Hodapp}, {Honda}, {Howell},
  {Hutsem{\'e}kers}, {Iono}, {Ip}, {Jackson}, {Jehin}, {Jiang}, {Jones},
  {Jones}, {Kadono}, {Kamath}, {K{\"a}ufl}, {Kasuga}, {Kawakita}, {Kelley},
  {Kerber}, {Kidger}, {Kinoshita}, {Knight}, {Lara}, {Larson}, {Lederer},
  {Lee}, {Levasseur-Regourd}, {Li}, {Li}, {Licandro}, {Lin}, {Lisse},
  {LoCurto}, {Lovell}, {Lowry}, {Lyke}, {Lynch}, {Ma}, {Magee-Sauer},
  {Maheswar}, {Manfroid}, {Marco}, {Martin}, {Melnick}, {Miller}, {Miyata},
  {Moriarty-Schieven}, {Moskovitz}, {Mueller}, {Mumma}, {Muneer}, {Neufeld},
  {Ootsubo}, {Osip}, {Pandea}, {Pantin}, {Paterno-Mahler}, {Patten},
  {Penprase}, {Peck}, {Petitpas}, {Pinilla-Alonso}, {Pittichova}, {Pompei},
  {Prabhu}, {Qi}, {Rao}, {Rauer}, {Reitsema}, {Rodgers}, {Rodriguez}, {Ruane},
  {Ruch}, {Rujopakarn}, {Sahu}, {Sako}, {Sakon}, {Samarasinha}, {Sarkissian},
  {Saviane}, {Schirmer}, {Schultz}, {Schulz}, {Seitzer}, {Sekiguchi}, {Selman},
  {Serra-Ricart}, {Sharp}, {Snell}, {Snodgrass}, {Stallard}, {Stecklein},
  {Sterken}, {St{\"u}we}, {Sugita}, {Sumner}, {Suntzeff}, {Swaters},
  {Takakuwa}, {Takato}, {Thomas-Osip}, {Thompson}, {Tokunaga}, {Tozzi}, {Tran},
  {Troy}, {Trujillo}, {Van Cleve}, {Vasundhara}, {Vazquez}, {Vilas},
  {Villanueva}, {von Braun}, {Vora}, {Wainscoat}, {Walsh}, {Watanabe},
  {Weaver}, {Weaver}, {Weiler}, {Weissman}, {Welsh}, {Wilner}, {Wolk},
  {Womack}, {Wooden}, {Woodney}, {Woodward}, {Wu}, {Wu}, {Yamashita}, {Yang},
  {Yang}, {Yokogawa}, {Zook}, {Zauderer}, {Zhao}, {Zhou}, \& {Zucconi}}]{maa05}
{Meech}, K.~J., {Ageorges}, N., {A'Hearn}, M.~F., {et~al.} 2005, Science, 310,
  265

\bibitem[{{Miyazaki} {et~al.}(2012){Miyazaki}, {Komiyama}, {Nakaya}, {Kamata},
  {Doi}, {Hamana}, {Karoji}, {Furusawa}, {Kawanomoto}, {Morokuma}, {Ishizuka},
  {Nariai}, {Tanaka}, {Uraguchi}, {Utsumi}, {Obuchi}, {Okura}, {Oguri},
  {Takata}, {Tomono}, {Kurakami}, {Namikawa}, {Usuda}, {Yamanoi}, {Terai},
  {Uekiyo}, {Yamada}, {Koike}, {Aihara}, {Fujimori}, {Mineo}, {Miyatake},
  {Yasuda}, {Nishizawa}, {Saito}, {Tanaka}, {Uchida}, {Katayama}, {Wang},
  {Chen}, {Lupton}, {Loomis}, {Bickerton}, {Price}, {Gunn}, {Suzuki},
  {Miyazaki}, {Muramatsu}, {Yamamoto}, {Endo}, {Ezaki}, {Itoh}, {Miwa},
  {Yokota}, {Matsuda}, {Ebinuma}, \& {Takeshi}}]{mkn12}
{Miyazaki}, S., {Komiyama}, Y., {Nakaya}, H., {et~al.} 2012, Proc. SPIE, 8446,
  0

\bibitem[{{Miyazaki} {et~al.}(2002){Miyazaki}, {Komiyama}, {Sekiguchi},
  {Okamura}, {Doi}, {Furusawa}, {Hamabe}, {Imi}, {Kimura}, {Nakata}, {Okada},
  {Ouchi}, {Shimasaku}, {Yagi}, \& {Yasuda}}]{mks02}
{Miyazaki}, S., {Komiyama}, Y., {Sekiguchi}, M., {et~al.} 2002, \pasj, 54, 833

\bibitem[{{Monet} {et~al.}(2003){Monet}, {Levine}, {Canzian}, {Ables}, {Bird},
  {Dahn}, {Guetter}, {Harris}, {Henden}, {Leggett}, {Levison}, {Luginbuhl},
  {Martini}, {Monet}, {Munn}, {Pier}, {Rhodes}, {Riepe}, {Sell}, {Stone},
  {Vrba}, {Walker}, {Westerhout}, {Brucato}, {Reid}, {Schoening}, {Hartley},
  {Read}, \& {Tritton}}]{mlc03}
{Monet}, D.~G., {Levine}, S.~E., {Canzian}, B., {et~al.} 2003, \aj, 125, 984

\bibitem[{{Mortara} \& {Fowler}(1981)}]{mof81}
{Mortara}, L. \& {Fowler}, A. 1981, Proc. SPIE, 290, 28

\bibitem[{{Nikolaev} {et~al.}(2000){Nikolaev}, {Weinberg}, {Skrutskie},
  {Cutri}, {Wheelock}, {Gizis}, \& {Howard}}]{nws00}
{Nikolaev}, S., {Weinberg}, M.~D., {Skrutskie}, M.~F., {et~al.} 2000, \aj, 120,
  3340

\bibitem[{{Onaka} {et~al.}(2008){Onaka}, {Tonry}, {Isani}, {Lee}, {Uyeshiro},
  {Rae}, {Robertson}, \& {Ching}}]{oti08}
{Onaka}, P., {Tonry}, J.~L., {Isani}, S., {et~al.} 2008, Proc. SPIE, 7014, 12

\bibitem[{{Pavelin} \& {Walter}(1980)}]{paw80}
{Pavelin}, C.~J. \& {Walter}, A.~J.~H. 1980, Proc. SPIE, 264, 70

\bibitem[{{Pe{\~n}a Ram{\'{\i}}rez} {et~al.}(2011){Pe{\~n}a Ram{\'{\i}}rez},
  {Zapatero Osorio}, {B{\'e}jar}, {Rebolo}, \& {Bihain}}]{pzb11}
{Pe{\~n}a Ram{\'{\i}}rez}, K., {Zapatero Osorio}, M.~R., {B{\'e}jar}, V.~J.~S.,
  {Rebolo}, R., \& {Bihain}, G. 2011, \aap, 532, A42

\bibitem[{{Perryman} \& {ESA}(1997)}]{esa97}
{Perryman}, M.~A.~C. \& {ESA}, eds. 1997, ESA Special Publication, Vol. 1200,
  {The HIPPARCOS and TYCHO catalogues. Astrometric and photometric star
  catalogues derived from the ESA HIPPARCOS Space Astrometry Mission}, ed.
  M.~A.~C. {Perryman} \& {ESA}

\bibitem[{{Persson} {et~al.}(1998){Persson}, {Murphy}, {Krzeminski}, {Roth}, \&
  {Rieke}}]{pmk98}
{Persson}, S.~E., {Murphy}, D.~C., {Krzeminski}, W., {Roth}, M., \& {Rieke},
  M.~J. 1998, \aj, 116, 2475

\bibitem[{{Roeser} {et~al.}(2010){Roeser}, {Demleitner}, \&
  {Schilbach}}]{rds10}
{Roeser}, S., {Demleitner}, M., \& {Schilbach}, E. 2010, \aj, 139, 2440

\bibitem[{{Santander-Garc{\'{\i}}a} {et~al.}(2008){Santander-Garc{\'{\i}}a},
  {Corradi}, {Mampaso}, {Morisset}, {Munari}, {Schirmer}, {Balick}, \&
  {Livio}}]{scm08}
{Santander-Garc{\'{\i}}a}, M., {Corradi}, R.~L.~M., {Mampaso}, A., {et~al.}
  2008, \aap, 485, 117

\bibitem[{Schirmer {et~al.}(2003)Schirmer, Erben, Schneider, Pietrzynski,
  Gieren, Micol, \& Pierfederici}]{ses03}
Schirmer, M., Erben, T., Schneider, P., {et~al.} 2003, A\&A, 407, 869

\bibitem[{{Schirmer} {et~al.}(2011){Schirmer}, {Hildebrandt}, {Kuijken}, \&
  {Erben}}]{shk11}
{Schirmer}, M., {Hildebrandt}, H., {Kuijken}, K., \& {Erben}, T. 2011, \aap,
  532, 57

\bibitem[{{Sekiguchi} {et~al.}(2002){Sekiguchi}, {Boehnhardt}, {Hainaut}, \&
  {Delahodde}}]{sbh02}
{Sekiguchi}, T., {Boehnhardt}, H., {Hainaut}, O.~R., \& {Delahodde}, C.~E.
  2002, \aap, 385, 281

\bibitem[{{Shupe} {et~al.}(2012){Shupe}, {Laher}, {Storrie-Lombardi}, {Surace},
  {Grillmair}, {Levitan}, \& {Sesar}}]{sls12}
{Shupe}, D.~L., {Laher}, R.~R., {Storrie-Lombardi}, L., {et~al.} 2012, Proc.
  SPIE, 8451, 1

\bibitem[{{Skrutskie} {et~al.}(2006){Skrutskie}, {Cutri}, {Stiening},
  {Weinberg}, {Schneider}, {Carpenter}, {Beichman}, {Capps}, {Chester},
  {Elias}, {Huchra}, {Liebert}, {Lonsdale}, {Monet}, {Price}, {Seitzer},
  {Jarrett}, {Kirkpatrick}, {Gizis}, {Howard}, {Evans}, {Fowler}, {Fullmer},
  {Hurt}, {Light}, {Kopan}, {Marsh}, {McCallon}, {Tam}, {Van Dyk}, \&
  {Wheelock}}]{scs06}
{Skrutskie}, M.~F., {Cutri}, R.~M., {Stiening}, R., {et~al.} 2006, \aj, 131,
  1163

\bibitem[{{Smith} {et~al.}(2007){Smith}, {Tucker}, {Allam}, {Ivezi{\'c}},
  {Yanny}, {Gunn}, {Knapp}, {Eisenstein}, {Finkbeiner}, \& {Fukugita}}]{sta07}
{Smith}, J.~A., {Tucker}, D.~L., {Allam}, S.~S., {et~al.} 2007, in ASP Conf.
  Ser., Vol. 364, The Future of Photometric, Spectrophotometric and
  Polarimetric Standardization, ed. C.~{Sterken}, 91

\bibitem[{{Stetson}(2000)}]{ste00}
{Stetson}, P.~B. 2000, \pasp, 112, 925

\bibitem[{{Tziamtzis} {et~al.}(2009){Tziamtzis}, {Schirmer}, {Lundqvist}, \&
  {Sollerman}}]{tsl09}
{Tziamtzis}, A., {Schirmer}, M., {Lundqvist}, P., \& {Sollerman}, J. 2009,
  \aap, 497, 167

\bibitem[{{Uchimoto} {et~al.}(2012){Uchimoto}, {Yamada}, {Kajisawa}, {Kubo},
  {Ichikawa}, {Matsuda}, {Akiyama}, {Hayashino}, {Konishi}, {Nishimura},
  {Omata}, {Suzuki}, {Tanaka}, {Tokoku}, \& {Yoshikawa}}]{uyk12}
{Uchimoto}, Y.~K., {Yamada}, T., {Kajisawa}, M., {et~al.} 2012, \apj, 750, 116

\bibitem[{{Valdes}(1984)}]{val84}
{Valdes}, F. 1984, in Bulletin of the American Astronomical Society, Vol.~16,
  Bulletin of the American Astronomical Society, 497

\bibitem[{{von der Linden} {et~al.}(2012){von der Linden}, {Allen},
  {Applegate}, {Kelly}, {Allen}, {Ebeling}, {Burchat}, {Burke}, {Donovan},
  {Morris}, {Blandford}, {Erben}, \& {Mantz}}]{laa12}
{von der Linden}, A., {Allen}, M.~T., {Applegate}, D.~E., {et~al.} 2012, ArXiv
  e-prints

\bibitem[{{Zacharias} {et~al.}(2012){Zacharias}, {Finch}, {Girard}, {Henden},
  {Bartlett}, {Monet}, \& {Zacharias}}]{zfg12}
{Zacharias}, N., {Finch}, C.~T., {Girard}, T.~M., {et~al.} 2012, VizieR Online
  Data Catalog, 1322, 0

\end{thebibliography}

\appendix{Online material}
\twocolumn

\section{\label{example1}Optical example: VLT/FORS1}
This section introduces the basic reduction steps with THELI, using a set of raw data taken from 
the ESO archive\footnote{{\tt http://archive.eso.org/eso/eso\_archive\_main.html}}. The idea is to provide 
the beginner with a simple data set and a sense for how THELI 
works. Only mandatory steps are covered, i.e.~preparation of the data, the main calibration, weighting, 
astrometry, sky subtraction and coaddition. The example is based on the small $R$-band FORS1@VLT data set 
listed in Table \ref{example1data} and taken by \cite{sbh02}, for TNO 1996 TO$_{66}$ (Fig.~\ref{tno}). At that 
time FORS1 was a single-chip camera. Treatment of multi-chip data is identical, with one exception 
occurring during astrometry (highlighted in App.~/ref{example2}).

\begin{table}[h]
\caption{Small FORS1@VLT data set for the example in App. \ref{example1}. Exposures can be identified in 
  the ESO raw science archive by selecting FORS1 as instrument, entering ``13 11 1999'' for the night, and by 
  limiting the maximum number of output rows to 200.}
\label{example1data}
\begin{tabular}{l}
\noalign{\smallskip}
\hline
\hline
\noalign{\smallskip}
File name \\
\hline
\noalign{\smallskip}
BIAS\\
{\tt FORS.1999-11-14T11:35:16.454.fits}\\
{\tt FORS.1999-11-14T11:36:02.607.fits}\\
{\tt FORS.1999-11-14T11:36:49.109.fits}\\
{\tt FORS.1999-11-14T11:37:35.510.fits}\\
{\tt FORS.1999-11-14T11:38:22.303.fits}\\
\hline
\noalign{\smallskip}
FLAT\\
{\tt FORS.1999-11-13T23:42:09.117.fits}\\
{\tt FORS.1999-11-13T23:43:21.663.fits}\\
{\tt FORS.1999-11-13T23:44:40.835.fits}\\
{\tt FORS.1999-11-14T09:17:20.006.fits}\\
{\tt FORS.1999-11-14T09:18:14.457.fits}\\
\hline
\noalign{\smallskip}
SCIENCE\\
{\tt FORS.1999-11-14T00:26:52.043.fits}\\
{\tt FORS.1999-11-14T00:38:56.322.fits}\\
{\tt FORS.1999-11-14T00:51:18.793.fits}\\
{\tt FORS.1999-11-14T01:03:58.970.fits}\\
{\tt FORS.1999-11-14T01:16:04.471.fits}\\
{\tt FORS.1999-11-14T01:28:27.545.fits}\\
{\tt FORS.1999-11-14T01:45:35.796.fits}\\
{\tt FORS.1999-11-14T01:57:40.224.fits}\\
{\tt FORS.1999-11-14T02:10:04.102.fits}\\
{\tt FORS.1999-11-14T02:27:48.341.fits}\\
\hline
\end{tabular}
\end{table}

\subsection{Initializing THELI and preparing the data}
Sort the uncompressed data into three arbitrarily named sub-directories {\tt BIAS}, {\tt FLAT} and 
{\tt SCIENCE}, which share the same parent directory, {\tt /MAINPATH}. Next, in THELI's \textit{Initialise} 
section, 
\begin{itemize}
\item{enter a project
name (e.g.~{\tt FORS1\_DEMO})}
\item{select the number of CPUs you want to use}
\item{select {\tt FORS1\_1CCD@VLT} from the instrument list}
\item{let THELI know the directory tree (Fig.~\ref{theliini}).}
\end{itemize}

In the \textit{Preparation} section, mark the \textit{Split FITS / correct header} task and launch it by 
clicking on \textit{Start}. THELI will cycle through the three data directories, split multi-extension FITS 
files into single FITS files (if applicable), and translates the FITS headers (Sect.~\ref{fixheader}). File 
names will end in {\tt \_i.fits}, where {\tt i} will run from 1 to $n$ for a multi-chip camera with $n$ 
detectors.

\subsection{Calibration}
Mark \textit{Process biases}, \textit{Process flats} and \textit{Calibrate data}. This creates the master 
bias and the bias-corrected master flat,\\
{\tt /MAINPATH/BIAS/BIAS\_1.fits},\\ 
{\tt /MAINPATH/FLAT/FLAT\_1.fits},\\
and applies both to the science data. All files are also overscan corrected and trimmed. The tasks can be 
executed individually or in one go. File names in {\tt SCIENCE} will now end in {\tt \_1OFC.fits}, where the 
status string {\tt OFC} indicates that the exposures have run through this pre-processing stage. Other steps 
at a later stage, such as background modeling, append more characters to the status string, but they are not 
required here.

\subsection{Weighting}
The global weight is a copy of the normalized flat field, where static bad pixels can optionally be zeroed 
(Sect.~\ref{weighting}). It forms the basis for the individual weight maps. All weight images can be found in 
{\tt /MAINDIR/WEIGHTS}.

Mark \textit{Create global weights} and \textit{Create WEIGHTs} and run both tasks with their default 
configuration settings. The global weight maps should be inspected for excessive masking if tighter 
lower and upper thresholds are chosen for the normalized flat.

\subsection{Astrometry and relative photometry}
First, we need to download the astrometric reference catalog. Select {\tt SDSS-DR9} and a magnitude limit of 23, 
so we get sufficiently many sources matching the VLT data. The search radius is automatically adjusted to $5^\prime$. 
Click on \textit{Get catalog} to download nearly 500 reference sources. Then run \textit{Create source cat} with its 
default parameters, which creates the binary catalogs for each exposure in {\tt /MAINPATH/SCIENCE/cat}. Therein
you will also find {\tt ds9cat} and {\tt skycat} sub-directories, containing catalogs formatted for overlay 
in these two FITS viewing programs. Two versions of the astrometric reference catalog, called 
{\tt ds9cat/theli\_mystd.reg} and {\tt skycat/theli\_mystd.skycat}, are kept there as well.

Now run the astrometry. Mark \textit{Astro+photometry}, select {\tt Scamp} from the pull-down menu next to
the task, and click on \textit{Configure}. The default settings work with a broad range of data. Click on 
\textit{Defaults (this page)} to make sure no different values from a previous run are loaded. Close the 
configuration dialog, and start the astrometry task. Once done, the sub-directory {\tt /MAINPATH/SCIENCE/plots} 
contains the {\tt Scamp} check plots which you should inspect\footnote{A large selection of good and bad {\tt Scamp} 
check plots is shown in the online user manual.}. The most important ones are

\begin{itemize}
\item{{\tt fgroups\_1.png}: The exposures' layout on sky. Green (red) marks indicate objects in the reference 
  catalog that were (not) matched by an object in the exposures. You should recognize the dither pattern. The 
  image frames displayed should not appear sheared or distorted, and you should recognize the dither pattern.}
\item{{\tt distort\_1.png}: The optical distortion, encoded as a change of pixel scale across the field. 
  This should be circularly symmetric and aligned with the optical axis of the telescope (usually at the center 
  of the detector array).}
\item{{\tt astr\_referror\_1.png}: The astrometric residuals with respect to the reference catalog. No 
  systematic trends should be visible, and the scattering should represent the astrometric uncertainties 
  of the reference catalog.}
\item{{\tt astr\_interror\_1.png}: The internal astrometric residuals, i.e.~the accuracy exposures were 
  registered with respect to each other. This should be on the order of 1/5th to 1/15th of a pixel, and 
  systematic trends should be absent.}
\end{itemize}

\subsection{Sky subtraction and coaddition}
No extended sources are present, therefore normal sky modeling is adequate. In the 
\textit{Coaddition} section, enter the configuration for sky subtraction, select \textit{Model the sky} 
and load the default parameters. Exit the dialog, mark and run the \textit{Sky subtraction} task. The 
names of the sky-subtracted images end in {\tt \_1OFC.sub.fits}.

Lastly, perform the image coaddition with the default parameter configuration. If you want, enter a proper 
motion vector of $\Delta {\rm RA} = -1.\!\!^{\prime\prime}531915\,{\rm hr}^{-1}$ and 
$\Delta {\rm DEC} = -0\,\myarcsec59283\,{\rm hr}^{-1}$ to register the exposures on the TNO, reproducing 
Fig.~\ref{tno}. Enter an extra \textit{Identification string} so that a previous coaddition without proper 
motion is not overwritten. Exit the configuration dialog and run the task. The final coadded image 
and weight can be found in {\tt SCIENCE/coadd\_<ID>/coadd.[weight].fits}, where 
{\tt <ID>} is either the filter in which the images were taken, or a user-supplied identifier.

\section{\label{example2}Near-IR example: HAWK-I@VLT}
This example is based on $H$-band observations of low mass objects in the $\sigma$ Orionis cluster
\citep{pzb11}. The necessary reduction steps for near-IR data are shown, focussing on the background 
modeling. The latter can be configured in many different ways, hence only the settings relevant for 
this data set are explained. Since HAWK-I@VLT consists of 4 HAWAII-2 detectors, this example also explains 
how astrometry is achieved for mosaic cameras. Settings already explained in App. \ref{example1} are 
not repeated here unless their importance is to be emphasized.

\subsection{Initializing THELI and preparing the data}
Sort the uncompressed data into sub-directories {\tt FLAT}, {\tt FLAT\_OFF} and {\tt SCIENCE}, sharing the  
same parent directory, {\tt MAINPATH}. Choose a new project name, e.g.~{\tt HAWKI\_DEMO}, select {\tt HAWKI@VLT} 
in the instrument list, and make the directory tree known to THELI.

In the \textit{Preparation} section, mark the \textit{Split FITS / correct header} task. This time file names
will end in {\tt \_i.fits}, where {\tt i} runs from 1 to 4, representing the four chips of HAWK-I.

\begin{table}[t]
\caption{HAWK-I@VLT data for the example in App.~\ref{example2}. The flats can be 
located in the ESO raw science archive by entering ``08 12 2008'' for the night, $H$ for the 
filter, and {\tt FLAT} for the data type. For the science exposures (abridged) choose
``07 12 2008'', select $H$ for the filter, and {\tt OBJECT} for the data type.}
\label{example2data}
\begin{tabular}{l}
\noalign{\smallskip}
\hline
\hline
\noalign{\smallskip}
File name \\
\hline
\noalign{\smallskip}
FLAT\\
{\tt HAWKI.2008-12-08T23:42:44.267.fits}\\ 
{\tt HAWKI.2008-12-08T23:42:56.627.fits}\\ 
{\tt HAWKI.2008-12-08T23:43:08.978.fits}\\ 
{\tt HAWKI.2008-12-08T23:43:21.381.fits}\\ 
{\tt HAWKI.2008-12-08T23:43:33.738.fits}\\ 
{\tt HAWKI.2008-12-08T23:43:46.669.fits}\\ 
{\tt HAWKI.2008-12-08T23:43:59.023.fits}\\ 
\hline
\noalign{\smallskip}
FLAT\_OFF\\
{\tt HAWKI.2008-12-08T23:52:04.524.fits}\\ 
{\tt HAWKI.2008-12-08T23:52:14.233.fits}\\ 
{\tt HAWKI.2008-12-08T23:52:26.606.fits}\\ 
{\tt HAWKI.2008-12-08T23:52:38.957.fits}\\ 
{\tt HAWKI.2008-12-08T23:52:51.921.fits}\\ 
{\tt HAWKI.2008-12-08T23:53:04.321.fits}\\ 
{\tt HAWKI.2008-12-08T23:53:17.273.fits}\\ 
\hline
\noalign{\smallskip}
SCIENCE\\
{\tt HAWKI.2008-12-08T04:40:22.699.fits}\\
{\tt HAWKI.2008-12-08T04:42:54.715.fits}\\
{\tt HAWKI.2008-12-08T04:45:28.151.fits}\\
{\tt HAWKI.2008-12-08T04:48:01.630.fits}\\
{[$\dots$]}\\
{\tt HAWKI.2008-12-08T05:44:19.241.fits}\\
{\tt HAWKI.2008-12-08T05:46:52.674.fits}\\
{\tt HAWKI.2008-12-08T05:49:26.152.fits}\\
{\tt HAWKI.2008-12-08T05:51:59.587.fits}\\
\hline
\end{tabular}
\end{table}

\subsection{Calibration}
First, set the \textit{Do not apply BIAS / DARK} check box as the pre-read is already subtracted from 
near-IR data\footnote{If you want to subtract a dark, remove this setting after the master
flat was created, and select \textit{Use dark} next to the \textit{Calibrate data} task.}. Run 
\textit{Process flats} to create combined bright and dark flats. The bright flat will have the dark flat 
subtracted automatically. If you do not want to subtract a dark flat, just do not provide that data.
Next, run \textit{Calibrate data} to apply the flat field to the science images.

\subsection{\label{2passexample}Background modeling}
Now a background model has to be removed, a task inevitable with near-IR data. This is done in the 
\textit{Background} 
section. First of all, we have to decide whether we want a one-pass or a two-pass background model. In most 
cases a two-pass approach is required, unless the field is empty, exposures were frequently and widely dithered, 
and / or accurate photometry is not necessary. Our field has several bright stars and may contain nebulosity. 
A two-pass strategy is therefore adequate, meaning the same task is run twice, the second time with fine-tuned 
parameters. 

\subsubsection{First pass of the background model}
Open the configuration for \textit{Background model correction}. You will be presented with the
following settings.
\begin{enumerate}
\item{{\bf Mask objects:} For the first pass, remove the default entries for {\tt DT} and {\tt DMIN}, 
  representing the {\tt SExtractor} {\tt DETECT\_THRESH} and {\tt DETECT\_MINAREA} parameters. This will create 
  a quick and simple background model without prior object masking\footnote{If you want object masking 
    at this stage, e.g.~because a two-pass is not necessary, choose {\tt DT} and {\tt DMIN} high enough such 
    that no background features will be masked, e.g.~{\tt DT=20, DMIN=10}.}. 
  Select the \textit{Median} combination, and switch \textit{off} the {\tt SExtractor} filtering.}
\item{{\bf Reject pixels from the stack:} By default, the highest pixel in the stack is rejected prior to 
  stack combination. Accept this setting. You may want to increase this number if you create larger static 
  stacks, have high source density, and / or larger dynamic window sizes.}
\item{{\bf How to apply the background model:} Different options are available. For near-IR data we 
  \textit{subtract} the model and leave the two smoothing scales empty. We want to \textit{rescale} the model
  to take out any temporal global intensity variations. We do not apply the model to SKY data (because we do 
  not have SKY exposures in this example), and we also do not adjust the gains between chips (done 
  during flat-fielding).}
\item{{\bf Static or dynamic model:} Choose a \textit{window size} of 6, and accept the 
  default value for the maximum gap size (1.0h; there is no gap in this exposure sequence if the 
  same exposures as listed in Table \ref{example2data} are used).}
\end{enumerate}

Once finished, the files will have the string {\tt OFCB} in their names. The previous {\tt OFC} images are 
parked in {\tt OFC\_IMAGES}, and a {\tt MASK\_IMAGES} sub-directory contains the object masks 
from which the background model was calculated. Since we chose to not detect objects, the masks are 
simply links to the unmasked data. The background models for each individual image can be found in the 
{\tt BACKGROUND} directory.

In case horizontal or vertical gradients from a reset anomaly are still present (as in Fig.~\ref{moircs}),
one should now run the \textit{Collapse correction} task (Sect.~\ref{collapse}). This is not necessary for 
the present data.

\subsubsection{Second pass of the background model}
At this point the images are mostly flat, revealing fainter objects (which can thus be masked). 
To create an improved background model we simply repeat the task with modified parameters {\tt DT=1.5} and 
{\tt DMIN=10}.

THELI will recognize that this is the second pass. Objects detected in the {\tt OFCB} images will be masked in 
the {\tt OFC} images (parked in {\tt OFC\_IMAGES}), and the improved masks can be found again under 
{\tt MASK\_IMAGES}. The temporary {\tt OFCB} images from the 1st pass will be moved to 
{\tt OFCB\_IMAGES\_1PASS}, and the original {\tt OFC} images are restored. Improved 
background models are calculated and applied, resulting in the final set of {\tt OFCB} images.

Inspect a few of the masks and the resulting {\tt OFCB} exposures, in particular a few at the beginning, the 
middle and at the end of the sequence. Relax {\tt DT=1.5} if excessive masking is recognized, and simply 
re-run the task. In general, two passes are sufficient. One without any detection thresholds (faster), 
and the second one with correctly chosen settings. If necessary, a collapse correction can be applied 
afterwards.

\subsubsection{Separate sky exposures}
If extended sources were present in these HAWK-I data, then the background model could not be calculated 
from the data themselves. Instead, separate exposures of a nearby blank field would be needed. They 
have to be collected in a {\tt SKY} directory (Sect.~\ref{skydata}), where THELI would find and automatically 
process them. This is not the case for this example, but I include it here as it is a common procedure. 

In the 
following, the letter 'O' denotes an object exposure, and the letter 'S' a sky exposure. For example, 
alternating sequences could be 
\begin{itemize}
\item{{\tt OOOOO-SSSSS}}
\item{{\tt OOO-SSS-OOO-SSS- ...}}
\item{{\tt OO-S-OO-S-OO-S- ...}}
\end{itemize}
The first one would be chosen for a short observation of a brighter target, justifying a static background model.
The other sequences could also be processed with a single static model, but the layout suggests that the 
observer wanted to keep track of the sky variations, calling for a dynamic model. The layout of the sequences 
\textit{does not matter} to THELI. You simply collect all sky exposures in a separate {\tt SKY} directory.  
For the dynamic model, THELI will identify the closest sky exposures in time for a given object exposure, 
based on the window size, and assuming that a valid (modified) Julian date {\tt MJD-OBS} is present in the FITS 
headers. The latter should be the case for all pre-defined instruments in THELI, but you may want to check, as 
the raw FITS headers may have changed or be corrupted.

\subsection{Weighting}
Weighting is essentially the same as for optical data. Tighter thresholds should be chosen for the normalized 
flat to remove spurious pixel clusters. Set the {\tt min} and {\tt max} values for {\tt FLAT\_norm} to 0.9 and 
1.1, respectively, and check the {\tt WEIGHT/globalweight\_i.fits} images for excessive masking.

\subsection{Astrometry and relative photometry}
For the reference catalog we choose again {\tt SDSS-DR9} and a magnitude limit of 23, which will result in about 
800 reference sources. For a field with high extinction 2MASS could be a better choice to maximize the overlap
between reference and source catalogs. \textit{Create source cat} can be run with its default parameters. For 
sparser fields than this example, {\tt DT} can be lowered from 5 to about 1.5 to get sufficiently many 
objects.

Use {\tt Scamp} for the astrometry, and select the default configuration parameters as for the first example.
However, set {\tt MOSAIC\_TYPE = SAME\_CRVAL} as we are processing data from a multi-chip camera. A focal plane 
(FP) model that has been determined previously based on observations of a dense stellar field will then form the 
basis 
of the final astrometric solution. This covers relative chip positions and rotations, as well as the 
CD-matrix\footnote{The prior CD-matrix will override the CD-matrix values from the FITS headers. If the position 
angle of your observations is different from the FP by more than {\tt POSANGLE\_MAXERR}, then the 
matching with the reference catalog will fail. Simply set {\tt POSANGLE\_MAXERR = 180} to allow for all 
position angles to be searched.}. Even sparse fields with only few detections per chip can thus be solved. A 
prior distortion model, however, cannot be loaded by current versions of {\tt Scamp}. The distortion is determined 
by {\tt Scamp} from the dithered data themselves. 

All pre-configured multi-chip instruments in THELI have their FPs already determined in this manner.
It may happen that these default FPs need to be updated, e.g.~after a dead detector has been
replaced, the optics re-aligned, or if the quality of the astrometric solution is insufficient. 
In these cases you can create a new FP from a different stellar field taken more recently, 
or simply from the current data set itself. To create a new FP, set {\tt FOCAL\_PLANE = Create new FP}, 
otherwise leave the default setting {\tt Use default FP}.

\subsection{Sky subtraction and coaddition}
In Sect.~\ref{2passexample} a dynamic background model was subtracted. Yet small constant non-zero offsets 
on the order of half a percent of the original background level may still be present in the images, together 
with spatial variations of similar amplitude if the atmosphere was particularly unstable. Running an individual 
sky background model as shown for the optical example suppresses these residuals to a level of $\sim0.1$\% or 
less. Alternatively, you can choose a constant sky subtraction.

The coaddition can be performed with default parameters.

\section{\label{example3}Mid-IR example: T-ReCS@GEMINI}
This example is based on multi-band observations of the circum-nuclear starburst ring in NGC 7552,
(Table \ref{example3}; PI B. Rodgers), and shown in Fig.~\ref{trecs-chopnod}. The data can be obtained from 
the Gemini Science Archive. T-ReCS is a highly sensitive mid-IR imaging spectrograph and was offered at 
Gemini-South until 2012. This example highlights the typical chop-nod process for ground-based 
mid-IR observations, the special requirements for astrometry, and THELI's general support for multi-color 
data sets during coaddition.

\begin{table}[t]
\caption{Multi-band T-ReCS@GEMINI data for the mid-IR example in Appendix \ref{example3}. 
The exposures can be found in the Gemini Science Archive, by entering {\tt NGC 7552} 
for the object, and {\tt T-ReCS} for the instrument.}
\label{example3data}
\begin{tabular}{ll}
\noalign{\smallskip}
\hline
\hline
\noalign{\smallskip}
File name & Bandpass\\
\hline
\noalign{\smallskip}
S20110722S0134.fits & [\ion{Si}{II}] 8.8$\mu$m \\
S20110722S0135.fits & [\ion{Si}{II}] 8.8$\mu$m \\
S20110722S0136.fits & [\ion{Ne}{II}] 12.8$\mu$m\\
S20110722S0137.fits & [\ion{Ne}{II}] 12.8$\mu$m\\
S20110722S0138.fits & [\ion{Ne}{II}] 13.1$\mu$m\\
S20110722S0139.fits & [\ion{Ne}{II}] 13.1$\mu$m\\
S20110726S0086.fits & Qa 18.3$\mu$m\\
S20110726S0087.fits & Qa 18.3$\mu$m\\
\hline
\end{tabular}
\end{table}

\subsection{Initializing THELI and preparing the data}
T-ReCS images (and those of other mid-IR cameras) are usually not flat fielded due to the highly 
variable mid-IR background. Likewise, a bias or dark subtraction is unnecessary as it is taken care of
automatically by the pairwise image subtraction. Therefore, all images listed in Table \ref{example3data}
may be copied into the same {\tt SCIENCE} directory. Choose a new project name, e.g.~{\tt NGC7552\_MIR}, 
select {\tt TRECS@GEMINI}, and make the directory tree known to THELI.

In the \textit{Preparation} section, mark the \textit{Split FITS / correct header} task, and check the 
\textit{Split MIR cubes} option. Contrary to 
optical and near-IR data, several additional processing steps take place for T-ReCS. First, THELI checks
whether the images were taken in full chop-nod mode, which is the case for almost all exposures with 
T-ReCS (and currently the only supported mode for this camera). The images used in this example typically 
contain $8-9$ extensions with a full chop-nod cycle. THELI will perform the chop-nod sky subtraction and 
write out separate images per cycle, as we asked to split the cubes. If the target was very faint and only 
visible in the stacked cube, then the splitting option should not be checked, as otherwise the astrometry 
will become unfeasible.

\subsection{Calibration}
As mentioned above, no dark or flat correction is done. Most of the sky and telescope 
background was already removed in the previous step. Nevertheless, the calibration task has to be run for
compatibility reasons, otherwise the data will not propagate properly through THELI. Simply set the 
\textit{Do not apply BIAS / DARK} and \textit{Do not apply FLAT} check boxes, and run \textit{Calibrate data}. 
The latter will merely insert the standard {\tt OFC} string into the file names, and set some internal 
processing flags.

\subsection{Collapse correction}
The chop-nod background correction for mid-IR depends strongly on the atmosphere's precipitable water vapor
and possible cirrus. Observations are only possible in dry and clear conditions, in particular in the $Q$ 
bandpass centered on 20$\mu$m. Chop-nod cycles taken in unstable conditions may show strong background 
residuals and should be discarded. This is why THELI does not automatically combine all cycles during 
splitting, but leaves them for individual inspection so that bad exposures can be discarded manually. 

For this example data set no images have to be rejected. However, you may notice that a small vertical gradient 
exists in some of the data. This can be easily corrected for using the \textit{Collapse correction}.
Adopt the default configuration values (leave {\tt DT} and {\tt DMIN} empty), with \textit{x} as the 
collapse correction. We want to exclude the entire extended and diffuse object from the modeling, defining 
the vertices of the excluded region as {\tt xmin=110}, {\tt xmax=220}, {\tt ymin=50}, and {\tt ymax=160}. This 
is for the [\ion{Si}{II}] 8.8$\mu$m, [\ion{Ne}{II}] 12.8$\mu$m, and [\ion{Ne}{II}] 13.1$\mu$m filters. The 
observations in the
Qa 18.3$\mu$m bandpass are significantly offset and require {\tt xmin=130}, {\tt xmax=240}, {\tt ymin=70}, and 
{\tt ymax=180}. To achieve this, park the Qa exposures in a temporary directory and run the collapse correction
with the first setting on the [\ion{Si}{II}]and [\ion{Ne}{II}] exposures. Then exchange the images in the 
current and the temporary directory, and apply the collapse correction to the Qa exposures with the updated 
exclusion region. After that, merge all images again in the {\tt SCIENCE} directory.

\subsection{Weighting}
As no flat field is available, the individual weights are based on constant global weights with 
value unity. Select \textit{Same weight for all pixels} in the configuration of \textit{Create global 
weights}. The individual weights can be created with default settings, i.e.~cosmics, hot pixels and obvious 
static pixel defects will be masked.

\subsection{Astrometry and photometry}
Mid-IR images probe a vastly different part of the spectrum than optical or near-IR images, and typically
only one or two compact sources or diffuse emission are seen. Matching with common astrometric reference
catalogs, a full distortion correction and/or WCS solution is in general not possible. Usually, a simple 
linear registration of images is sufficient. The global WCS in the final coadded image will be limited by 
the accuracy of the WCS in the FITS raw data, and may have to be adjusted by hand if necessary.

Two possibilities exist for mid-IR data. In case of a single point source, the simple \textit{Shift (float)}
approach can be used, requiring the creation of source catalogs as discussed in the previous examples. This 
approach does not work here, as one or several point sources may be detected in the individual images, 
together with extended flux. The number of sources is very small, and hence an unambiguous cross-identification 
of exposures is unlikely. Imagine you have one image with one source detected, and another one with two sources. 
Without further prior information, such as assumptions about the accuracy of the WCS in the header, it is not 
possible to register these two images automatically. Use the cross-correlation method (\textit{Xcorr}) instead. 
It will run {\tt SExtractor} with fixed low detection thresholds to create images containing compact and extended 
source flux only. These are then cross-correlated, yielding the relative offsets.

\subsection{\label{prepcolor}Sky subtraction and coaddition}
After \textit{collapse correction} the mean sky background level is zero, therefore sky subtraction can be 
skipped. If the collapse correction was not done, the images may still show some constant offset which can be 
removed using the \textit{Mode} in \textit{Subtract a constant sky}.

Coaddition is performed with default settings. However, since we have observations in 4 different bands 
present in the same directory, we want the final coadded images to have identical geometries. We also want the 
same physical object to appear in the same pixel. This is achieved as follows:

\begin{enumerate}
\item{First, find the RA and DEC values of the center of your image. It does not need to be accurate, but if 
  you are too far off, the coadded image could be truncated. Enter these values as {\tt Ref RA|DEC} in the 
  configuration of the \textit{Coaddition} dialog. For this example we use {\tt RA=23:16:10.6} and 
  {\tt DEC=-42:35:04}. By using the same values for the coadditions of all 4 bands, we enforce an identical 
  astrometric deprojection needed for automatic image registration later on.}
\item{\textit{Coadd this filter:} In this pull-down menu THELI presents a list of the 4 bands available in the 
  {\tt SCIENCE} directory. Start with the first one, {\tt NeII-12.8um}, which will create a 
  {\tt coadd\_NeII-12.8um} directory containing the corresponding coadded image.}
\item{Optionally, you can choose an additional outlier rejection by setting {\tt threshold=3} and 
  {\tt cluster size=4}}
\end{enumerate}

At this point, four {\tt coadd\_<filter>} directories are present, with the stacked images and their weights, 
{\tt coadd.[weight].fits}. While these have identical values of {\tt CRVAL} and {\tt CRPIX}, their geometries 
are still different due to initial pointing variations, chop-nod offsets etc. To register the images, go to the 
main menu at the top of the THELI window, and open \textit{Miscellaneous $\rightarrow$ Prepare color picture}. 
You will be presented with a list of all {\tt coadd\_<filter>} directories. Select the ones you are
interested in, and then click on \textit{Get coadded images}. This will create a {\tt SCIENCE/color\_theli}
directory, containing the coadded images registered and trimmed to their maximum common overlap, and called
{\tt <filter>\_cropped.[weight].fits}. Note that since THELI enforces integer values for the {\tt CRPIX}
header keywords during coaddition, excess pixels can simply be removed and no second resampling has to take 
place that would degrade image quality.

This approach works for any other multi-color data set. The only requirement is that identical reference 
RA and DEC values are used for the coadditions of the different bands, and that the final {\tt coadd\_<ID>}
directories are collected in the same {\tt SCIENCE} directory.

\section{\label{insttables}Currently supported instruments}

\begin{table*}
\caption{\label{opticalinsts}Pre-configured optical imagers}
\begin{tabular}{lll}
  \noalign{\smallskip}
  \hline 
  \hline 
  \noalign{\smallskip}
  Instrument@Telescope & \# detectors & Comment\\
  \noalign{\smallskip}
  \hline 
  \noalign{\smallskip}
  ACAM@WHT			    &  1  & \\
  ALFOSC@NOT			    &  1  & \\
  ALTAU16M@VYSOS06		    &  1  & \\
  AltaU42@ASV		            &  1  & Low and high resolution modes\\
  CFH12K@CFHT			    &  12 & Before and after detector swap in September 1999\\
  DECam@CTIO			    &  62 & \\
  EFOSC2@ESO3.6m		    &  1  & $1\times1$ and $2\times2$ binning\\
  EMMI@NTT		            &  1  & BIMG and RILD imaging modes\\
  ENZIAN\_CAS@HOLI\_1M		    &  1  & \\
  FORS1@VLT		            &  1  & old and new configuration with 1 and 2 CCDs, respectively\\
  FORS2@VLT		            &  1  & old and new configuration with 1 and 2 CCDs, respectively\\
  GMOS@GEMINI-NORTH		    &  3  & $1\times1$ and $2\times2$ binning\\
  GMOS@GEMINI-SOUTH		    &  3  & $1\times1$ and $2\times2$ binning\\
  GOODMAN@SOAR			    &  1  & $1\times1$ and $2\times2$ binning\\
  GPC1@PS1      		    &  64 & \\
  GROND\_OIMG@MPGESO		    &  1  & \\
  IMACS\_F2@LCO		            &  8  & old and new detector configuration\\
  IMACS\_F4@LCO		            &  8  & old and new detector configuration\\
  LAICA\_2x2@CAHA		    &  4  & $1\times1$ and $2\times2$ binning\\
  LBC@LBT			    &  4  & BLUE and RED cameras\\
  LDSS3@LCO			    &  2  & \\
  LORRI@NewHorizons		    &  1  & \\
  MEGACAM@LCO     		    &  36 & $2\times2$ binning only\\
  MEGAPRIME@CFHT		    &  36 & Support for raw and ELIXIR pre-processed data\\
  MEROPE@MERCATOR		    &  1  & \\
  MOSAIC-I@KPNO\_0.9m	            &  8  & Old configuration (before Aug. 2010) \\
  MOSAIC-I@KPNO\_4.0m	            &  8  & Old configuration (before Aug. 2010) \\
  MOSAIC-II@CTIO                    &  8  & 8- and 16-channel mode, before and after fix of dead readout port\\
  MOSCA\_2x2@NOT		    &  4  & $2\times2$ binning only\\
  OASIS4x4@WHT			    &  1  & $1\times1$ and $4\times4$ binning\\
  OMEGACAM@VST			    &  32 & \\
  PFC@WHT			    &  2  & \\
  SDSS                              &  1  & For SDSS images directly downloaded from the server\\
  SOI@SOAR			    &  2  & \\
  SuprimeCam\_OLD@SUBARU            &  10 & Before and after April 2001 (replacement of dead CCD)\\
  SuprimeCam\_NEW@SUBARU            &  10 & Installed August 2008; Support for raw and SDFRED pre-processed data\\
  SuSI2\_2x2@NTT		    &  2  & $2\times2$ binning only\\
  VIMOS@VLT			    &  4  & \\
  WFC@INT			    &  4  & $1\times1$ and $2\times2$ binning\\
  WFI@AAT			    &  8  & \\
  WFI@SSO\_40inch		    &  7  & \\
  WFI@MPGESO			    &  8  & \\
  Y4KCam@CTIO			    &  1  & \\
  \hline 
  \noalign{\smallskip}
\end{tabular}
\end{table*}

\begin{table*}
\caption{\label{irinsts}Pre-configured near- and mid-IR imagers}
\begin{tabular}{lll}
  \noalign{\smallskip}
  \hline 
  \hline 
  \noalign{\smallskip}
  Instrument@Telescope & \# detectors & Comment\\
  \noalign{\smallskip}
  \hline 
  \noalign{\smallskip}
  FLAMINGOS2@GEMINI-SOUTH 	     & 1  & \\
  FourStar@LCO                       & 4  & \\
  GROND\_IRIM@MPGESO		     & 1  & \\
  GSAOI@GEMINI-SOUTH 	      	     & 4  & \\
  HAWKI@VLT			     & 4  & \\
  INGRID@WHT			     & 1  & \\
  ISAAC@VLT			     & 1  & \\
  LIRIS@WHT		             & 1  & Normal imaging and polarimetry mode\\
  MMIRS@LCO			     & 1  & \\
  MOIRCS@SUBARU		             & 4  & \\
  NACOSDI@VLT			     & 1  & \\
  NEWFIRM@CTIO			     & 4  & \\
  NICS@TNG			     & 1  & \\
  NICI@GEMINI-SOUTH		     & 2  & \\
  NIRI@GEMINI-NORTH		     & 1  & \\
  NOTcam@NOT		             & 1  & High and low resolution modes\\
  Omega2000@CAHA		     & 1  & \\
  OSIRIS@SOAR		             & 1  & High and low resolution modes\\
  OSIRIS@GTC			     & 1  & \\
  PISCES@LBT          		     & 1  & \\
  SOFI@NTT			     & 1  & \\
  SPARTAN@SOAR			     & 4  & \\
  TRECS@GEMINI			     & 4  & mid-infrared\\
  VIRCAM@VISTA			     & 16 & \\ 
  VISIR@VLT			     & 1  & mid-infrared\\
  WIRCam@CFHT			     & 4  & \\
  \hline
  \noalign{\smallskip}
\end{tabular}
\end{table*}

\end{document}